\documentclass[letterpaper,onecolumn,11pt,accepted=2021-12-25]{quantumarticle}
\pdfoutput=1
\usepackage[utf8]{inputenc}
\usepackage[english]{babel}
\usepackage[T1]{fontenc}
\usepackage{amsmath}
\usepackage{hyperref}

\usepackage{tikz}
\usepackage{lipsum}

\numberwithin{equation}{section}

\newcounter{resultcounter}[section]

\newtheorem{thm}[resultcounter]{Theorem}
\newtheorem{lem}[resultcounter]{Lemma}
\newtheorem{prop}[resultcounter]{Proposition}

\newcommand{\e}{\,{\rm e}}

\newcommand{\aes}{a_e^{(s)}}
\newcommand{\Qes}{Q_e^{(s)}}

\newcommand{\Ees}{E_e^{(s)}(\lambda)}

\usepackage{amsmath}
\usepackage{amssymb}
\usepackage{graphicx}
\usepackage{color}

\renewcommand{\r}{{\rm R}}
\newcommand{\s}{{\rm S}}

\renewcommand{\i}{{\rm i}}

\def\qed{\hfill $\Box$\medskip}

\newcommand{\bbbone}{\mathchoice {\rm 1\mskip-4mu l} {\rm 1\mskip-4mu l}
	{\rm 1\mskip-4.5mu l} {\rm 1\mskip-5mu l}}

\begin{document}
	
\title{Dynamics of 	
		Open Quantum Systems II,  
		
		Markovian Approximation}	
\author{Marco Merkli}
\affiliation{Department of Mathematics and Statistics, Memorial University of Newfoundland, St. \!John's, A1C 5S7, Canada}
\orcid{0000-0002-3990-6155}
	\email{merkli@mun.ca}
\homepage{https://www.math.mun.ca/~merkli/}
\maketitle

\centerline{\small Dedication: {\em Per Bap.}}

\begin{abstract}
A finite-dimensional quantum system is coupled to a bath of oscillators in thermal equilibrium at temperature $T>0$. We show that for fixed, small values of the coupling constant $\lambda$, the true reduced dynamics of the system is approximated by the completely positive, trace preserving markovian semigroup generated by the Davies-Lindblad generator. The difference between the true and the markovian dynamics is $O(|\lambda|^{1/4})$  for all times, meaning  that the solution of the Gorini-Kossakowski-Sudarshan-Lindblad  master equation is approximating the true dynamics to accuracy $O(|\lambda|^{1/4})$ for all times. Our method is based on a recently obtained expansion of the full system-bath propagator. It applies to reservoirs with correlation functions decaying in time as slowly as $1/t^{4}$, which is a significant  improvement relative to the  previously required exponential decay.
\end{abstract}

\section{Introduction}

\subsection{Motivation}

The fundamental evolution equation of quantum theory is the Schr\"odinger equation
\begin{equation}
\i \frac{d}{dt} \psi(t) = H\psi(t),
\end{equation}
where $\psi(t)$ is a pure state (normalized vector) in a Hilbert space $\mathcal H$ and $H$ is the Hamiltonian, the operator representing the energy observable. (We choose units in which we have effectively $\hbar=1$.) Equivalently, density matrices $\rho(t)$, which are mixtures of pure states, evolve according to the von-Neumann equation 
\begin{equation}
\i \frac{d}{dt}\rho(t) = [H,\rho(t)].
\end{equation}
The state of a part, or subsystem, of a whole system, is obtained by `tracing out' the degrees of freedom of the complement. This is formalized as follows. The Hilbert space is bipartite, ${\mathcal H}={\mathcal H}_\s\otimes{\mathcal H}_\r$ ($\s$ corresponding to the subsystem one is interested in and $\r$ being the remaining part) and the {\em reduced state of $\s$} is 
\begin{equation}
\rho_\s(t)={\rm tr}_\r \, \rho(t) = {\rm tr}_\r \big( e^{-\i t H}\rho(0) e^{\i t H}\big),
\end{equation}
where ${\rm tr}_\r$ is the partial trace with respect to ${\mathcal H}_\r$. What is the evolution equation for $\rho_\s(t)$? Generically, when $\s$ and $\r$ interact, which is expressed by the fact that $H$ contains an interaction operator, the dynamics $\rho_\s(t)$ is very complicated, as it contains all information about the dynamics of $\r$ and the influence of $\r$ on $\s$. Except in trivial situations, it is impossible to find a system operator $H_{\rm eff}$, s.t. $\rho_\s(t)$ would equal $e^{-\i t H_{\rm eff}}\rho_\s(0)e^{\i t H_{\rm eff}}$.\footnote{Such an operator $H_{\rm eff}$, when it exists, could be called an `effective system Hamiltonian', even though it might not be hermitian. The associated `dynamics' $e^{-\i t H_{\rm eff}}\rho_\s(0)e^{\i t H_{\rm eff}}\equiv e^{t{\mathcal L}}\rho_\s(0)$ is called {\em unitary} if $H_{\rm eff}$ is hermitian, and it is called {\em markovian} if $H_{\rm eff}$ is hermitian or not.} Indeed, because $\s$ and $\r$ become entangled during the evolution (as they interact), the assignment $t\mapsto \rho_\s(t)$ is not markovian, by which we mean that if we denote $\rho_\s(t)=V_t\rho_\s(0)$, where $V_t$ is the `flow operator' of the evolution, then $V_{s+t}\neq V_sV_t$. Equality here would hold if $V_t$ was generated by an $H_{\rm eff}$, as mentioned above.

An important task in the study of open quantum systems is to identify under what circumstances  the true reduced  dynamical map $V_t$ can be {\em approximated} as
\begin{equation}
V_t \approx e^{t\mathcal L},
\end{equation}
by a (semi-)group $\e^{t{\mathcal L}}$, where $\mathcal L$ is a `superoperator' acting on density matrices and playing the role of an effective Hamiltonian (or, the operator taking the commutator with an effective Hamiltonian).\footnote{By an `effective Hamiltonian', we mean a generally non-hermitian generator of dynamics -- non-hermiticity  is natural here since the open system dynamics will generically exhibit irreversible effects, such as decoherence, thermalization...}  Analyzing the approximating dynamics $e^{t{\mathcal L}}$ can then be done by standard means, namely by diagonalizing $\mathcal L$ and expressing the propagator $e^{t\mathcal L}$ as $\sum_j e^{t \epsilon_j}P_j$ with eigenvalues $\epsilon_j$ and eigenprojections $P_j$. The approximate dynamics is expressed equivalently as
\begin{equation}
\frac{d}{dt}\rho_{\s,{\rm M}}(t) = {\mathcal L}\rho_{\s,{\rm M}}(t),
\end{equation}
where the subscript ``$\rm M$'' stands for {\em markovian}.  The latter equation is called the markovian master equation. Very often, one is interested in the situation where $\r$ is a `very large' quantum system, also called a `reservoir'. Then it is expected that the reduced system dynamics shows irreversibility. For instance, if $\r$ is a thermal bath (a large number of quantum oscillators in thermal equilibrium) at a given temperature then one expects that $\rho_{\s, \rm M}(t)$ converges to an equilibrium state at the same temperature, when $t\rightarrow\infty$.  Irreversibility of the system dynamics is then described by negative real parts ${\rm Re}\, \epsilon_j<0$, of the `effective energies' $\epsilon_j$ which are the eigenvalues of $\mathcal L$. 

Deriving a markovian effective evolution equation for open systems is an important task for theoretical and practical reasons \cite{AL,BreuerPetruccione,Lidarnotes,RivasHuelga}. The literature on the topic is huge and growing, and it encompasses research in mathematics, physics, chemistry, quantum information and biology  \cite{AN,BF,CP,C,FP,Lidar,MajLid,Moh,ML, R1,R2}.  In typical heuristic derivations, one explains that the markovian approximation should be valid under two main conditions \cite{BreuerPetruccione,RivasHuelga,AL, Lidarnotes}:
\begin{itemize}
	\item  The reservoir correlation time (decay time of its correlation function) should be much shorter than the system relaxation time (typical time by which the open system $\s$ changes significantly). This means that the reservoir loses its memory quickly relative to the pace at which the system moves, so information is carried away from the system and does not flow back to it, which effectively erases memory in the dynamics and makes it markovian.
	\item The reservoir state, initially in equilibrium, does not significantly change during the course of time. This is plausible if the  interaction between $\s$ and $\r$ is weak and the reservoir is much bigger than the system $\s$. 
\end{itemize}

There are numerous papers deriving candidates for the generator $\mathcal L$ obtained by making a small coupling expansion of the dynamics (weak $\s\r$ interaction, gauged by an interaction constant $\lambda$). This is done both in general and specific open system models, and the forms and properties of the proposed $\mathcal L$ vary. Over the course of time, the so-called ``Davies generator'' has emerged as the correct, or ``proper'' generator \cite{DS}. The (semi-)group $e^{t\mathcal L}$ generated by the Davies generator is CPTP (completely positive, trace preserving), guaranteeing that under its evolution, an initial system density matrix stays a density matrix at all times -- other generators do not have this property (but they still might approximate the true density matrix, at least on some restricted time-scales). The justification of the validity of the markovian approximation generated by the Davies generator was given in \cite{Davies1,Davies2}. There it was shown that for time scales up to $O(1/\lambda^2)$, where $\lambda$ is the $\s\r$ interaction constant, the difference between the true and the approximate dynamics vanishes in the limit $\lambda \rightarrow 0$. However, in any derivation up to and including \cite{Davies1,Davies2}, the question of whether the markovian approximation is valid for all times (beyond $O(1/\lambda^2)$) could not be answered with certainty. In particular, there were speculations as to what the correct asymptotic state is for the system (as $t\rightarrow\infty$). A detailed account of this point is given in the recent survey \cite{TMCA}. Over the last decade, the so-called quantum resonance theory was able to show that the markovian approximation (given by the Davies generator) is valid for all times \cite{MAOP,KM2}. 

Our goal here is to improve those results and make them applicable for a larger class of models. In \cite{MAOP,KM2}, a crucial assumption was that the reservoir correlations decay exponentially quickly in time. This is equivalent to having an interaction operator which is very regular (namely, analytic; see Section 2.6 of \cite{MAOP} for a detailed discussion). Here we develop a method which can deal with less regular interactions, equivalent to only polynomially decaying reservoir correlations (as slow as $\sim t^{-4}$). We explain the details in Section \ref{sect:model}. It is not a mere technical exercise to see how much regularity, or correlation decay, is needed to guarantee the markovian approximation to be valid. Namely, it is known that eventually, when the interactions are too `rough', the markovian approximation does {\em not} hold \cite{R1,R2,Li}.

\bigskip
\bigskip

\subsection{Model}
\label{sect:model}

Consider an $N$-level system with Hilbert space ${\mathbb C}^N$  coupled to a reservoir of oscillators in equilibrium at temperature $T>0$.  The Hamiltonian is given by
\begin{equation}
H_\lambda = H_\s +H_\r +\lambda G\otimes\varphi(g),
\label{n1}
\end{equation}
where $H_\s$ is an $N\times N$ hermitian matrix with (possibly degenerate) eigenvalues $E_j$ and eigenvectors $\phi_j$,
\begin{equation}
\label{n2}
H_\s = \sum_{j=1}^N E_j |\phi_j\rangle\langle\phi_j|
\end{equation}
and $H_\r$ is the environment, or reservoir Hamiltonian
\begin{eqnarray}
\label{n3}
H_\r = \sum_{k} \omega_k a^*_ka_k,
\end{eqnarray}
describing modes of a collection of harmonic oscillators, labelled by $k$, with frequencies  $\omega_k>0$ (put $\hbar=1$). The creation and annihilation operators $a^*_k$, $a_k$ satisfy the canonical commutation relations $[a_k,a^*_\ell] = \delta_{k,\ell}$ (Kronecker symbol). The interaction term contains a coupling constant $\lambda\in\mathbb R$, an interaction operator $G$ (hermitian $N\times N$ matrix), and it is linear in the field operator 
\begin{equation}
\label{n4}
\varphi(g) = \frac{1}{\sqrt 2}\sum_{k} g_k a^*_k + {\rm h.c.},
\end{equation}
where ${\rm h.c.}$ denotes the hermitian conjugate.  The collection of the numbers $g_k\in \mathbb C$ constitutes the form factor $g$. The size of $g_k$  determines how strongly the oscillator $k$ is coupled to the system. This model is ubiquitous, for $N=2$ it is called the spin-Boson model.

In the limit of continuous modes (thermodynamic limit, infinite volume limit...), the values of frequencies $\omega_k$ becomes a continuum, as do the values of the parameter $k$.  For definiteness, we consider a reservoir modeling a (scalar) quantized field in three dimensions. This corresponds to momenta $k\in{\mathbb R}^3$, and $\omega_k$, $g_k$, $a^*_k$ and $a_k$ become functions $\omega(k)$, $g(k)$, $a^*(k)$, $a(k)$ with $[a(k),a^*(\ell)] = \delta(k-\ell)$ (Dirac distribution).  In the continuous mode limit, the reservoir Hamiltonian \eqref{n3} and field operator \eqref{n4}  are
\begin{eqnarray}
H_\r &=& \int_{{\mathbb R}^3} \omega(k) a^*(k)a(k)d^3k,\label{resham}\\
\varphi(g) &= & \frac{1}{\sqrt 2}\int_{{\mathbb R}^3} \big(g(k) a^*(k) +{\rm h.c.}\big) d^3k. \label{n5}
\end{eqnarray}
We assume for convenience that $\omega(k)=|k|$. The Hilbert space on which the operators \eqref{n5} act is the customary Bosonic Fock space over the single particle wave function space $L^2({\mathbb R}^3,d^3k)$ (momentum representation), 
\begin{equation}
\label{36.f}
{\mathcal F} = \oplus_{n\ge 0} \, L_{\rm sym}^2({\mathbb R}^{3n}, d^{3n}k), 
\end{equation}
where the subscript sym refers to symmetric functions  (Bosons) and the summand with $n=0$ is interpreted to be $\mathbb C$.

{\em In this paper, we understand that the continuous mode limit is performed to begin with, that is, we consider the Hamiltonian \eqref{n1} with $H_\s$ given in \eqref{n2} and $H_\r$, $\varphi(g)$ given in \eqref{n5}.} The Hilbert space of pure states of the coupled system is 
\begin{equation}
{\mathfrak h} ={\mathbb C}^N\otimes{\mathcal F}
\label{n43}
\end{equation}
with ${\mathcal F}$ given by \eqref{36.f}. 

We consider initially disentangled system-servoir states of the form
\begin{equation}
\rho_{\s\r,0} = \rho_\s\otimes\omega_{\r,\beta},
\label{n8}
\end{equation} 
where $\rho_\s$ is an arbitrary system ($N\times N$) density matrix and $\omega_{\r,\beta}$ is the equilibrium state of the reservoir at temperature $T=1/\beta>0$. It is not necessary to take initial product states as in \eqref{n8} for our method to work -- we study the case of correlated (classically or quantum) initial $\s\r$ states in \cite{Markov3}. Since the Hamiltonian $H_\r$ has continuous spectrum, the operator $e^{-\beta H_\r}$ is not trace class and so we cannot define the reservoir Gibbs density matrix directly. Rather, one performs a thermodynamic limit: By constraining the reservoir particles to move in a finite volume $\Lambda\subset {\mathbb R}^3$ in (typically a box in direct, position) space, the momenta of particles are quantized and discrete and one can define the usual equilibrium Gibbs density matrix $\rho^\Lambda_{\r,\beta}\propto \e^{-\beta H^\Lambda_\r}$ for $H^\Lambda_\r$ given in \eqref{n3} and frequencies determined by $\Lambda$. In the limit $|\Lambda|\rightarrow\infty$ of contiunous modes the `density matrix' $\propto e^{-\beta H^\Lambda_\r}$ is ill defined. However, the two-point function
$$
\omega_{\r,\beta}^\Lambda(a^*(f)a(g))=\frac{{\rm tr}\big( e^{-\beta H_\r^\Lambda} a^*(f)a(g)\big)}{{\rm tr}\ e^{-\beta H_\r^\Lambda}} \quad \longrightarrow \quad \omega_{\r,\beta}\big(a^*(f) a(g)\big),\qquad |\Lambda|\rightarrow\infty
$$ 
has a well defined limit, which can be explicitly evaluated as
\begin{equation}
\label{n31}
\omega_{\r,\beta}\big( a^*(f) a(g) \big) = \langle g,(\e^{\beta \omega(k)}-1)^{-1} f\rangle =\int_{{\mathbb R}^3} \frac{\bar g(k)f(k)}{\e^{\beta\omega(k)}-1} d^3k,
\end{equation}
for any $f,g\in L^2({\mathbb R}^3,d^3k)$ (where the overbar $\bar{\ }$ denotes complex conjugation here). The state $\omega_{\r,\beta}$, {\em a priori} defined only by \eqref{n31}, is extended to arbitrary products and sums of creation and annihilator operators by linearity and  (Gaussian) quasi-freeness, meaning that averages of general {\em field (reservoir) observables $\mathcal P$}, which are polynomials of creation and annihilation operators and limits thereof, are found using Wick's theorem \cite{BR}. This state is the infinite volume, or continuous mode limit of the reservoir in thermal equilibrium, and it is the initial state for the reservoir in \eqref{n8}. 

The state $\omega_{\r,\beta}$  is a linear functional on the observable algebra ${\mathcal P}$ of polynomials in creation and annihilation operators, or the Weyl algebra, and it can also be represented by a density matrix $\rho_{\r,\beta}$ (having actually rank one), albeit in a Hilbert space different from ${\mathcal F}$. This is called a purification and we explain it in Section \ref{s2.1} below. It is not too surprising that the Fock space $\mathcal F$, \eqref{36.f},  is not appropriate to describe a state with nonzero particle density in infinite volume, because every vector in $\mathcal F$ obtained by applying an arbitrary (finite) number of creation operators to the vacuum represents a state with finitely many particles, which means a particle density of {\em zero}, in the infinite volume limit.  The reduced system density matrix $\rho_\s(t)$ is defined by the relation
\begin{equation}
{\rm tr}_{{\mathbb C}^N} \big(\rho_\s(t) A\big) =\rho_\s\otimes\omega_{\r,\beta}\big(e^{\i t H_\lambda} (A\otimes\bbbone_\r) e^{-\i t H_\lambda}\big)
\label{10}
\end{equation}
valid for all system observables $A\in{\mathcal B}({\mathbb C}^N)$ (operators on ${\mathbb C}^N$). On the right side of \eqref{10}, we use the convenient notation $\rho_\s\otimes\omega_{\r,\beta}(A\otimes B) = {\rm tr}_{{\mathbb C}^N}(\rho_\s A) \omega_{\r,\beta}(B)$, which extends by linearity to all (even non-product) system-reservoir observables. The mapping $\rho_\s\mapsto \rho_\s(t)$ is linear and defines the so-called {\em dynamical map}
\begin{equation}
V_t : \rho_\s\mapsto V_t\rho_\s = \rho_\s(t).
\label{11}
\end{equation}
$V_t$ maps the initial system state to the state at time $t$, evolved according to the interacting dynamics.

A central question in the theory of open quantum systems is when one can approximate $V_t$ by a {\em completely positive trace preserving} (CPTP) semi-group $e^{t{\mathcal L}}$. A classical result is that this can be done under some conditions: The works \cite{Davies1, Davies2, VanHove} show that 
\begin{equation}
\lim_{\lambda\rightarrow 0} \sup_{0\le \lambda^2 t<a} \|V_t -e^{t({\mathcal L}_\s+\lambda^2K)}\| =0\qquad \mbox{for any $a>0$}.
\label{12}
\end{equation}
Here, ${\mathcal L}_\s=-\i [H_\s,\cdot]$ is the free system dynamics and $K$ describes the influence of the bath to second order in perturbation, consisting of a Hamiltonian as well as a dissipative term (see also \cite{AL,BreuerPetruccione,RivasHuelga}). Often, the superoperator $K$ is called the {\em Davies generator} and ${\mathcal L}_\s+\lambda^2K$ is the generator of a CPTP semigroup \cite{CP,DS,MAOP}. The Davies generator can be calculated by perturbation theory (and corresponds to the so-called level shift operator, see Section \ref{s2.3} and also the Appendix of \cite{MAOP}).  As follows from \eqref{12}, it can also be constructed from the relation
\begin{equation}
\lim_{\lambda\rightarrow 0}\ V_{\tau/\lambda^2}\, \circ\, e^{-\tau {\mathcal L}_\s/\lambda^2} = e^{\tau K},\qquad \tau>0.
\end{equation}
In differential form, the approximate markovian dynamics, defined as
\begin{equation}
\rho_{\rm M}(t) = e^{t({\mathcal L}_\s+\lambda^2K)}\rho_{\rm M}(0),
\label{12.1}
\end{equation} 
takes the form of the ubiquitous {\em markovian master equation}
\begin{equation}
\frac{d}{dt} \rho_{\rm M}(t) = {\mathcal L}\rho_{\rm M}(t), \qquad {\mathcal L}={\mathcal L}_\s+\lambda^2K.
\label{13}
\end{equation}
An inconvenience of \eqref{12} is that the accuracy of the markovian approximation is only guaranteed for times which are not too large, satisfying $\lambda^2 t < a$. This holds for arbitrary fixed $a$, but the deviation of the true dynamics $V_t$ from its markovian approximation $e^{t({\mathcal L}_\s+\lambda^2K)}$ cannot be bounded uniformly in $a$ -- in other words, the speed of the convergence to zero in \eqref{12}, as $\lambda\rightarrow 0$, depends on $a$. Said differently, \eqref{12} shows that  the markovian approximation is guaranteed to hold for $t\rightarrow\infty$ only if at the same time, one takes $\lambda\rightarrow 0$ with a bounded value for $\lambda^2 t$.  For this reason,  \eqref{12} is sometimes  called the weak-coupling (or van Hove) limit. Recently, in \cite{KM2,MAOP},  the result \eqref{12} was improved to 
\begin{equation}
\sup_{t\ge 0}\big \| V_t -e^{t({\mathcal L}_\s+\lambda^2K)}\big\| \le C \lambda^2, 
\label{14}
\end{equation}
provided $|\lambda|\le \lambda_0$ for some $\lambda_0$. As \eqref{14} shows, the markovian approximation is now guaranteed to be accurate to $O(\lambda^2)$ {\em uniformly in time} $t\ge 0$, for fixed $\lambda$. The proof of \eqref{14} is given using the so-called dynamical resonance theory. In \cite{KM2,MAOP}, an analyticity condition on the form factor $g(k)$ (see \eqref{n1}) is imposed, which  implies that the symmetrized reservoir correlation function 
\begin{equation}
C_\beta(t) = {\rm Re}\, c_\beta(t)\label{16.1},
\end{equation}
where 
\begin{equation}
c_\beta(t) = \omega_{\r,\beta}\big(\varphi(g) e^{\i t H_\r}\varphi(g)e^{-\i t H_\r}\big)
\label{16.2}
\end{equation}
is the correlation function, 
decays exponentially quickly in time, for large times. It is well known that decay of reservoir correlations is necessary for a markovian description of the system dynamics to be accurate, but demanding exponential decay is not necessary. The reason for the analyticity condition, and hence the assumption of exponential decay of correlations made in \cite{KM2,MAOP}, is that the analysis is based on so-called spectral deformation methods (pioneered in \cite{BFS,JP} in this setting). However, a technically more demanding approach, based on Mourre theory, was introduced in the current context in \cite{Merkli2001,FM} and further developed in \cite{KoMeSo,KM1}. It can be viewed as an infinitesimal version of the deformation method and only requires  the existence of {\em a few derivatives} (up to order four for us here), rather than {\em analyticity} of the form factor $g(k)$. Consequently, the Mourre theory approach allows us to deal with polynomially decaying reservoir correlation functions. In the present work, the decay has to be at least as quick as
\begin{equation}
C_\beta(t) \sim t^{-4}, \qquad t\rightarrow\infty,
\label{18}
\end{equation}
as we explain below,  after \eqref{98}.
\medskip

The {\bf main achievement} of the present work is a proof of the accuracy of the markovian approximation uniformly in time, assuming the above-mentioned weak regularity condition. Our main result is Theorem \ref{thm1}, which gives \eqref{14} but, alas, with an inferior error estmate of $O(|\lambda|^{1/4})$.  Prior to the current work, Mourre theory was not used to show the validity of the markovian approximation. In \cite{KoMeSo} it was used to show that the spin-Boson model has the property of return to equilibrium even in the strong coupling regime (in which one is forced to use Mourre theory as the analytic deformation cannot be applied). In \cite{KM1}, that approach was further developed to examine the reduced dynamics of the spin-Boson model (large coupling), giving an expansion of the spin dynamics into the sum of a main term plus a remainder. The remainder was shown to decay to zero as $t\rightarrow\infty$ but it was not shown to be small in $\lambda$, nor was it shown that the main term is a CPTP dynamics.  In \cite{Markov1} an expansion of the system-reservoir dynamics was derived with an explicitly identified main term (in terms of resonance data) and a remainder which is small in $\lambda$ for all times $t\ge 0$. Our current work uses the results of \cite{Markov1} to prove the validity of the markovian approximation.

\subsection{Main result}

Our main result is Theorem \ref{thm1} below. We first present the assumptions and discuss them right below. Recall that the full Hamiltonian is given in \eqref{n1}, with the reservoir Hamiltonian \eqref{resham} and the field operator $\varphi(g)$ as in \eqref{n5}, and $g=g(k)$ is called the form factor. 
\bigskip

\noindent
{\bf Assumptions\ }
\begin{itemize}
	\item[(1)] {\em Smoothness of the form factor} $g(k)\in L^2({\mathbb R}^3,d^3k)$.
	
	We assume  
	\begin{equation}
	g(k) = \frac{|k|^p}{1+|k|^{p+q}} h(k),
	\label{gcond}
	\end{equation}
	for some values 
	\begin{equation}
	\qquad p=-\tfrac12, \tfrac12, \tfrac32 \mbox{\ or\ } p>2\mbox{\quad and\quad} q>2,
	\label{paramvalues}
	\end{equation}
	and where $h(k)$ is a function four times differentiable with respect to $|k|$ for $|k|>0$,  such that $h$ and its derivatives are bounded on $(0,\infty)$ and $\lim_{|k|\rightarrow 0}\partial^j_{|k|}h(k)\neq 0$, $j=0,\ldots,4$. 
	
	\item[(2)] {\em Effective coupling}:
	\begin{itemize} 
		\item[(a)] Fermi Golden Rule Condition. We assume that 
		\begin{equation}
		\min_{\{1\le m,n\le N : E_m\neq E_n\}} \big|\langle\phi_m, G\phi_n\rangle\big|^2 \int_{S^2} \big|g\big(|E_m-E_n|,\Sigma\big)\big|^2d\Sigma \, >\, 0,
		\label{n97}
		\end{equation}
		where the minimum is over all distinct eigenvalue pairs of $H_\s$, see \eqref{n2}. Here, $S^2$ is the unit sphere in momentum space ${\mathbb R}^3$ and the function $g$ is represented in spherical coordinates $k=(|k|,\Sigma)$, $|k|\ge 0$ (length), $\Sigma\in S^2$ (angles).
		
		\item[(b)] Simplicity of resonance energies. We assume that the so-called {\em level shift operators} $\Lambda_e$, $e\in{\mathcal E}_0=\{ E_m-E_n\ :\ m,n=1,\ldots, N\}$, have simple eigenvalues. The $\Lambda_e$ are explicit matrices describing the second order ($\lambda^2$) corrections to the energy differences $e\in{\mathcal E}_0$, given in Section \ref{FGRsect}.
	\end{itemize}
\end{itemize}

\smallskip

\noindent
{\bf Discussion of  the assumptions\ }
\begin{itemize}
	\item[(i)] In the open system literature, precise conditions on the form factor (coupling function) $g(k)$ are often not presented. Rather, only the so-called reservoir spectral density,
	\begin{equation}
	J(\omega) = \tfrac{1}{2}\pi \omega^2\int_{S^2} |g(\omega,\Sigma)|^2 d\Sigma,\qquad \omega\ge 0,
	\label{specdens}
	\end{equation}
	is specified. This is so because many physical quantities, such as decay rates, depend on $g$ only in the form of $J$, see Section \ref{subsub:spinboson}. However, to derive those results rigorously, one will need direct assumptions on $g$. This is why Assumption (1) is phrased in terms of $g$ and not $J$. This assumption says that  $g$ should be four times differentiable in the radial variable (energy) $|k|$ and that it should have the infrared and ultraviolet behaviours $g(k)\sim |k|^p$ for $|k|\rightarrow 0$ and $g(k)\sim |k|^{-q}$ or $|k|\rightarrow\infty$, respectively, with $p$ and $q$ satisfying \eqref{paramvalues}. 
	
	In terms of the spectral density, condition (1) means that we assume  that $J(\omega)$ is a function which can be differentiated four times in $\omega > 0$  and that $J(\omega)\sim \omega^{2p+2}$ for $\omega\sim 0$ and $J(\omega)\sim \omega^{2-2q}$ for $\omega\sim \infty$, with $p$ and $q$ satisfying \eqref{paramvalues}.
	
	The spectral density is called sub-Ohmic, Ohmic or super-Ohmic according to whether  $J(\omega)\sim \omega^s$ for $\omega\sim 0$, with $s<1$, $s=1$ or $s>1$, respectively.  In terms of our parameter $p$ in \eqref{gcond}, \eqref{paramvalues}, we have  $s=2(p+1)=1,3,5$ and $s>6$. This means that we assume to be in the Ohmic or the super-Ohmic regime. 
	

	\item[(ii)] The inequality \eqref{n97} is  equivalent to 
	$$
	J(|E_m-E_n|) \neq 0\qquad\mbox{and\qquad $\langle\phi_m, G\phi_n\rangle\neq 0$,}\qquad\mbox{ for all $E_m\neq E_n$.}
	$$
	Physically, this means that the spectral density must not vanish at any Bohr energy $|E_m-E_n|$ of the system, and that the transition matrix elements of the interaction operator $G$ must not vanish between different energy eigenstates $\phi_m$ of the system. These are standard second-order perturbation theory assumptions, they guarantee that $O(\lambda^2)$ energy exchange processes betweeen $\s$ and $\r$ are not suppressed.

	Assumptions (1) together with (2a) guarantee that for small nonzero $\lambda$, the coupled system-reservoir complex has a unique stationary state, namely the coupled equilibrium state (see \cite{Merkli2001} and also Section \ref{FGRsect}).

	Assumption (2b) is a simplification that can be quite easily removed by a slightly more cumbersome analysis. We do not present it here. The assumption means that the generator of the effective system evolution (the so-called level shift operators) does not have degenerate eigenvalues. The condition is generically satisfied even in rather complex systems, see for instance \cite{MBR}. 
\end{itemize}
\bigskip

\noindent
Here is the main result of our paper:

\begin{thm}[Markovian approximation uniform in time.] 
	\label{thm1}
	There is a constant $c_0>0$ such that if  $|\lambda|< c_0$, then 
	\begin{equation}
	\|V_t -e^{t ({\mathcal L}_\s+\lambda^2K)} \|\le C|\lambda|^{1/4},
	\label{16}
	\end{equation}
	for another constant $C$. Here, ${\mathcal L}_\s=-\i [H_\s,\cdot]$ and $K$ is the Davies generator. 
\end{thm}
The norm in \eqref{16} is that of superoperators, {\em i.e.}, operators acting on density matrices. In particular, for any initial system density matrix $\rho_\s$, we have $\|V_t\rho_\s-e^{t\mathcal L}\rho_\s\|_1\le C|\lambda|^{1/4}$ (trace norm) with $C$ independent of $\rho_\s$.\footnote{Since we are in finite dimensions, all norms are actually equivalent.}

\medskip

The next result gives an estimate of $c_0$ in terms of the temperature, for low temperatures. 

\begin{prop}
	\label{prop1.2}
	Consider temperatures $T=1/\beta$ in the range $0<T<T_0$, for a fixed $T_0$. Then the constant $c_0$ in Theorem \ref{thm1} satisfies
	\begin{equation}
	c_0 =  c'_0\, T^{24}\, \Big(\frac{\min\big[ 1,a,\delta/\kappa^2, g^{3/2}\big]}{\max\big[ 1,\alpha,\kappa(1+\kappa/\delta)\big]}\Big)^{4/3}, 
	\label{lambdasmall}
	\end{equation}
	where $c'_0$ is a constant depending only on $T_0$, but independent of $\lambda$, $T$ and independent of the parameters $a,\alpha,\delta,\kappa,g$. The latter five quantities naturally emerging from the perturbation theory, some of which depend on $T$. They are given in \eqref{FGR}-\eqref{g} and they are explicitly calculable in concrete models (see Proposition \ref{prop1.3} below).
\end{prop}
The bound \eqref{lambdasmall} is not optimal. Finding the smallest possible $c_0$ for which our method still works is an arduous task which we do not take on at this stage.

\subsubsection{ Example: The spin-Boson model}
\label{subsub:spinboson}

As an explicit illustration, we consider the two-dimensional system with Hamiltonian $H_\s$ and interaction operator $G$, written in an orthonormal basis $\{\phi_1= |\!\uparrow\rangle, \phi_2=|\!\downarrow\rangle\}$, to be given by
\begin{equation} 
H_\s=\frac{\Delta}{2}
\begin{pmatrix}
1 & 0\\
0 & -1
\end{pmatrix},\qquad 
G=
\begin{pmatrix}
G_{11}&  G_{12}\\
\overline{G}_{12} & G_{22}
\end{pmatrix}.
\label{131.1}
\end{equation}
Here, $\Delta>0$ is the Bohr energy of the system and $G=G^*$ is a hermitian matrix. We introduce the superoperator $K$, acting on $2\times2$ density matrices $\rho$ as
\begin{eqnarray}
K \rho  &=& -\tfrac12 \widehat c_\beta(0) \big(G_{11}-G_{22}\big)^2\big( p_1\rho p_2 +p_2\rho p_1\big)\nonumber\\
&&+ \widehat c_\beta(\Delta)|G_{12}|^2\big(\rho_{11} p_2- \tfrac12  p_1\rho - \tfrac12 \rho p_1\big)+\widehat c_\beta(-\Delta) |G_{12}|^2\big(\rho_{22} p_1- \tfrac12  p_2\rho - \tfrac12 \rho p_2\big)\nonumber\\
&& -\i [H_{\rm LS},\rho]
\label{K}
\end{eqnarray}
with 
\begin{equation}
H_{\rm LS} = \frac{1}{\pi}\sum_{k,\ell=1,2} |G_{k\ell}|^2  \Big({\rm P.V.} \int_{\mathbb R} \frac{\widehat c_\beta(u)}{E_k-E_\ell-u}\Big) p_k.
\label{K'}
\end{equation}
In \eqref{K}, \eqref{K'}, we use the notation $p_j=|\phi_j\rangle\langle \phi_j|$, $\rho_{ij}=\langle \phi_i,\rho\phi_j\rangle$ for $j=1,2$,  $E_1=\Delta/2$, $E_2=-E_1$, and 
$$
\widehat c_\beta(u) = \frac{J(|u|)}{|1-e^{-\beta u}|},\qquad u\in{\mathbb R}.
$$ 
We understand $\widehat c_\beta(0)$ as the limit $\lim_{\omega\rightarrow 0}\widehat c_\beta(\omega)$ (which is nonzero only for $p=1/2$, {\em c.f.} \eqref{paramvalues}). 

The super-operators $-\i [H_{\rm LS},\cdot]$ and ${\mathcal L}_\s=-\i [H_\s,\cdot]$ commute.  Set 
\begin{equation}
{\mathcal L}={\mathcal L}_\s+\lambda^2 K.
\label{1.31}
\end{equation} 
It is readily verified that if $G_{12}\neq 0$ and $\widehat c_\beta(\Delta)\neq 0$ ({\em i.e.}, if \eqref{n97} holds), then the only density matrix $\rho$ satisfying ${\mathcal L}\rho=0$ is the Gibbs equilibrium,
\begin{equation}
\rho_{\s,\beta} = \frac{\e^{-\beta H_\s}}{{\rm tr} \,e^{-\beta H_\s}},
\end{equation}
and that $\lim_{t\rightarrow\infty} e^{t{\mathcal L}}\rho=\rho_{\s,\beta}$ for any initial state $\rho$. By integrating $\frac{d}{dt}\rho = {\mathcal L}\rho$, one easily finds explicitly,
\begin{eqnarray}
\big(e^{t{\mathcal L}}\rho\big)_{11} &=& e^{-\lambda^2\gamma_1 t} \big(\rho(0)\big)_{11} -\Big(1- e^{-\lambda^2\gamma_1 t}\Big)\, (\rho_{\s,\beta})_{11}\\
\big(e^{t{\mathcal L}}\rho\big)_{12} &=&e^{-\i t (\Delta +\lambda^2x_2)} e^{-\lambda^2\gamma_2 t}\big(\rho(0)\big)_{12}
\end{eqnarray}
with
\begin{eqnarray}
\gamma_1 &=& |G_{12}|^2 J(\Delta) \coth\big(\beta\Delta/2\big)\\
\gamma_2 &=& \tfrac12 \gamma_1 + \tfrac12 \widehat c_\beta(0)\big(G_{11}-G_{22}\big)^2\\
x_2&=&\tfrac1\pi \big( (G_{22})^2-(G_{11})^2\big) {\rm P.V.} \int_{{\mathbb R}}\frac{\widehat c_\beta(u)}{u}du \nonumber\\
&& + \tfrac1\pi |G_{12}|^2\,  {\rm P.V. } \int_{{\mathbb R}} J(|u|)\frac{\coth(\beta u/2)}{\Delta-u}du.
\end{eqnarray}
The  constants $\lambda^2\gamma_1$ and $\lambda^2\gamma_2$ are the thermalization and decoherence rates, respectively. 

\begin{prop}
	\label{prop1.3}
	There is a constant $c_0>0$ such that if $|\lambda|<c_0$, then \eqref{16} holds for ${\mathcal L}_\s = -\i[H_\s,\cdot]$ and $K$ given in \eqref{K}. Moreover, if $0<T<T_0$ for some fixed $T_0$, and if $p>-1/2$ (see \eqref{paramvalues}), then 
	\begin{equation}
	c_0=c'_0\, T^{76/3},
	\end{equation} 
	for a $c'_0$ depending only on $T_0$ and $\Delta$.
\end{prop}

Our estimates are not uniform in the temperature $T$, as $T\rightarrow 0$. One might attempt to optimize the bound $T^{76/3}$ given in Proposition \ref{prop1.3}, but we are not dealing with this here.

\section{Proof of Theorem \ref{thm1}}

The dynamics of an open quantum systems is usually presented in form of the evolution $t\mapsto \rho(t)$ of the density matrix. The link to our result, Corollary 3.2 given in 
 \cite{Markov1}, is made by passing to a purification of the system, that is, by representing the density matrix $\rho(t)$ by a {\em vector} in a new Hilbert space. (After all, a density matrix is an operator, which is an element -- or vector -- of the space of all linear operators.) We present the purification in Section \ref{s2.1} and in Section \ref{s2.2} we explain how the dynamics is expressed in the new Hilbert space. We show in Section \ref{s2.3} how the setup is precisely that of \cite{Markov1} and we how we can borrow the result of Corollary 3.2 there to prove Theorem \ref{thm1} here.

\subsection{Purification of the initial state}
\label{s2.1}

We now describe the purification of the system-reservoir complex. Take any system density matrix $\rho_\s$ acting on ${\mathfrak h}_\s$ and write it in its diagonal form, $\rho_\s=\sum_j p_j |\psi_j\rangle\langle\psi_j|$. Then associate to $\rho_\s$ the vector $\psi_\s=\sum_j\sqrt{p_j}\,  \psi_j\otimes{\mathcal C}\psi_j\in{\mathbb C}^N\otimes{\mathbb C}^N$. Here, $\mathcal C$ is the anti-linear operator taking complex conjugates of the components of a vector in ${\mathbb C}^N$ in the energy basis $\{\phi_j\}$, \eqref{n2} (the choice of this basis is a convention). The map $\mathcal C$ is introduced so that $\rho_\s\mapsto \Psi_\s$ is linear.  Also, define the linear map $\pi_\s$ (a so-called $*$-representation) taking matrices on ${\mathbb C}^N$ to matrices on ${\mathbb C}^N\otimes {\mathbb C}^N$ determined by 
\begin{equation}
A\mapsto \pi_\s(A)=A\otimes\bbbone_\s.
\label{n44.2}
\end{equation} 
It is then immediately verified that 
\begin{eqnarray}
{\rm tr}_{{\mathbb C}^N} \big(\rho_\s A\big) &=& \langle\Psi_\s,\pi_\s(A)\Psi_\s\rangle_{{\mathcal H}_\s},\label{n44.1}\\
{\mathcal H}_\s &=& {\mathbb C}^N\otimes {\mathbb C}^N.
\label{n44}
\end{eqnarray}
Relation \eqref{n44} is the purification of the state $\rho_\s$; it is also called the Gelfand-Naimark-Segal (GNS) representation. As an example, the equilibrium (Gibbs) state is represented by
\begin{equation} 
\rho_{\s,\beta} = Z_{\s,\beta}^{-1} \e^{-\beta H_\s} \quad \mapsto \quad \Omega_{\s,\beta} = Z_{\s,\beta}^{-1/2} \sum_{j=1}^N e^{-\beta E_j/2}\phi_j\otimes\phi_j,
\label{n47.1}
\end{equation}
where $Z_{\s,\beta}={\rm tr}_{{\mathbb C}^N} e^{-\beta H_\s}= \sum_{j=1}^N e^{-\beta E_j}$ is the partition function.  
\medskip

The purification of the reservoir equilibrium state is known as the {\em Araki-Woods representation} of the canonical commtation relations \cite{AW,MLnotes}. The starting point is the reservoir equilibrium state $\rho_{\r,\beta}\propto \e^{-\beta H_\r}$ for $H_\r$ given in \eqref{n3}. In the contiuous mode limit $e^{-\beta H_\r}$ is not trace class since $H_\r$ has continuous spectrum, so the `density matrix' $\propto e^{-\beta H_\r}$ is ill defined. Rather, the state is defined to be the continuous (thermodynamic) limit of a sequence of discrete modes equilibrium states. The limiting state is described by the normalized linear functional determined by the two-point function \eqref{n31}. The state is quasi-free, or Gaussian, meaning that averages of general field observables $\mathcal P$ (polynomials of creation and annihilation operators and limits thereof) are found using Wick's theorem. Let $\mathcal P$ be a field observable. Similarly to \eqref{n44}, we have a representation (which we verify to be correct in \eqref{n55} below)
\begin{equation} 
\omega_{\r,\beta} ({\mathcal P}) = \langle \Omega_\r, \pi_\beta({\mathcal P})\Omega_\r\rangle_{{\mathcal H}_\r}
\label{n47}
\end{equation}
where 
\begin{eqnarray}
{\mathcal H}_\r &=& \bigoplus_{n\ge 0} L^2_{\rm sym}\big( ({\mathbb R}\times S^2)^n, (du\times d\Sigma)^n\big)
\label{n48}
\end{eqnarray}
is the (symmetric) Fock space over the one-particle space 
\begin{equation}
L^2( {\mathbb R}\times S^2, du\times d\Sigma)\equiv L^2( {\mathbb R}\times S^2).
\end{equation} 
Here, $d\Sigma$ is the uniform measure over the unit sphere $S^2\subset {\mathbb R}^3$. The vector $\Omega_\r$ is the vacuum vector of the Fock space ${\mathcal H}_\r$, {\em i.e.}, $\Omega_\r=(1,0,0,\ldots)$ when written according to the sum decomposition \eqref{n48}. As in any Fock space, we have  creation and annihilation operators which we denote by $a_\beta^*(f)$, $a_\beta(f)$ where $f\in L^2({\mathbb R}\times S^2)$ is a single-particle function over which the Fock space ${\mathcal H}_\r$ is built. The canonical commutation relations $[a_\beta(f), a^*_\beta(g)] = \langle f,g\rangle_{L^2({\mathbb R}\times S^2)}$ are satisfied. The field operators acting on ${\mathcal H}_\r$ are defined by
\begin{equation} 
\varphi_\beta(f)=\frac{1}{\sqrt 2}\big(a_\beta^*(f) +a_\beta(f)\big).
\end{equation} 
As in any Fock space the annihilators map the vacuum vector to zero,  $a_\beta(f)\Omega_\r=0$ for all $f$. 

The representation map $\pi_\beta$ maps operators on ${\mathcal F}$, \eqref{36.f}, to operators on ${\mathcal H}_\r$. $\pi_\beta$ is linear and satsifies $\pi_\beta(A^*)=[\pi_\beta(A)]^*$. Its action on field operators $\varphi(g) =\frac{1}{\sqrt 2}(a^*(g)+a(g))$, where $g\in L^2({\mathbb R}^3,d^3k)$ is a one-particle state associated to the Fock space \eqref{36.f}, then determines $\pi_\beta$ uniquely.\footnote{\label{footnote2}For instance, as $a^*(g) = \frac{1}{\sqrt 2}(\varphi(g)-\i\varphi(\i g))$, we have $\pi_\beta(a^*(g))=\frac{1}{\sqrt 2}[\pi_\beta(\varphi(g))-\i\ \pi_\beta(\varphi(\i g))]$. 
} We have
\begin{equation}
\pi_\beta(\varphi(g)) = \varphi_\beta(\tau_\beta g),
\label{n48.1}
\end{equation}
where 
\begin{equation}
\tau_\beta : L^2({\mathbb R}^3, d^3k)\rightarrow  L^2( {\mathbb R}\times S^2, du\times d\Sigma)
\end{equation}
takes a function $g(k)$, $k\in{\mathbb R}^3$, into a function $(\tau_\beta g)(u,\Sigma)$, $u\in\mathbb R$, $\Sigma\in S^2$, defined by
\begin{equation}
\big(\tau_\beta g\big)(u,\Sigma) = \sqrt{\frac{u}{1-\e^{-\beta u}}} \   |u|^{1/2}\left\{
\begin{array}{ll}
g(u,\Sigma), & u\ge0\\
-\bar g(-u,\Sigma) & u<0
\end{array}
\right. .
\label{n50}
\end{equation}
On the right side of \eqref{n50}, $g$ is represented in spherical coordinates, $u=|k|\ge 0$, $\Sigma\in S^2$. (For instance, a radially symmetric form factor $g(k)=|k|^\alpha e^{-|k|/\kappa}$, $k\in{\mathbb R}^3$, is given in spherical coordinates by  $g(u,\Sigma)=u^\alpha e^{-u/\kappa}$, $u\ge 0$, independent of $\Sigma$.) Note hat the map $\tau_\beta$ {\em is not linear} (but only real linear), as on the second line in \eqref{n50} the complex conjugate $\bar g$ appears. The linearity of $\pi_\beta$ and \eqref{n48.1} imply that  $\pi_\beta(a^*(g))=\frac{1}{\sqrt 2}(\varphi_\beta(g_\beta)-\i\varphi_\beta(\{\i g\}_\beta)$ (see also Footnote \ref{footnote2}). Using \eqref{n50} then gives 
\begin{equation}
\pi_\beta(a^*(f)) = a_\beta^*\Big(\textstyle\sqrt{\frac{u}{1-\e^{-\beta u}}}\ |u|^{1/2} f(u,\Sigma)\chi_+(u) \Big) - a_\beta\Big(\textstyle\sqrt{\frac{u}{1-\e^{-\beta u}}}\ |u|^{1/2} \bar f(-u,\Sigma)\chi_-(u) \Big), 
\end{equation}
where $\chi_+(u)=1$ if $u\ge 0$ and $\chi_+(u)=0$ for $u<0$ (and $\chi_-(u)=1-\chi_+(u)$).

To verify that \eqref{n47} is correct with our definitions \eqref{n48}, \eqref{n48.1} and \eqref{n50}, we calculate for $f,g\in L^2({\mathbb R}^3,d^3k)$, 
\begin{eqnarray}
\langle \Omega_\r, \pi_\beta(a^*(f)a(g))\Omega_\r\rangle_{{\mathcal H}_\r} &=&\langle \Omega_\r, a_\beta\Big(\textstyle\sqrt{\frac{u}{1-\e^{-\beta u}}}\ |u|^{1/2} \bar f(-u,\Sigma)\chi_-(u) \Big) \nonumber \\
&&\ \ \times a^*_\beta\Big(\textstyle\sqrt{\frac{u}{1-\e^{-\beta u}}}\ |u|^{1/2} \bar g(-u,\Sigma)\chi_-(u) \Big)\Omega_\r \rangle\nonumber\\
&=&  
\int_{{\mathbb R}\times S^2}  \ \big| \frac{u}{1-\e^{-\beta u}}\big| |u| f(-u,\Sigma)\bar g(-u,\Sigma)\chi_-(u) du d\Sigma\nonumber\\
&=&\int_{{\mathbb R}_+\times S^2}\frac{u^2}{\e^{\beta u}-1} \bar g(u,\Sigma) f(u,\Sigma)du d\Sigma\nonumber\\
&=& \int_{{\mathbb R}^3} \frac{\bar g(k) f(k)}{\e^{\beta |k|}-1} d^3k,
\label{n55}
\end{eqnarray}
which reproduces \eqref{n31} correctly ($\omega(k)=|k|$). 
\medskip

{\bf Upshot of the purification procedure.}  Let us consider initial states of the disentangled form 
\begin{equation}
\rho_\s\otimes\omega_{\r,\beta},
\label{n56}
\end{equation} 
where $\rho_\s$ is an arbitrary density matrix acting on ${\mathbb C}^N$ and $\omega_{\r,\beta}$ is the reservoir equilibrium state defined by \eqref{n31}. In the Hilbert space 
\begin{equation}
{\mathcal H} = {\mathcal H}_\s\otimes{\mathcal H}_\r,
\end{equation}
where ${\mathcal H}_\s$ and ${\mathcal H}_\r$ are given in \eqref{n44} and \eqref{n48}, respectively, the state \eqref{n56} is represented by the normalized vector
\begin{equation}
\Psi_0=\Psi_\s\otimes\Omega_\r \in {\mathcal H}.
\label{n60}
\end{equation}
Here, $\Psi_\s\in{\mathcal H}_\s$ is the purification of $\rho_\s$. Moreover, 
\begin{equation}
\pi=\pi_\s\otimes\pi_\beta
\end{equation} 
(see \eqref{n44.2} and \eqref{n48.1}) maps observables $\mathcal O$ acting on ${\mathbb C}^N\otimes{\mathcal F}$ (before purification) to operators on ${\mathcal H}$ (after purification), and we have 
\begin{equation}
\big(\rho_\s\otimes\omega_{\r,\beta}\big)({\mathcal O}) = \langle \Psi_0, \pi({\mathcal O})\Psi_0\rangle_{\mathcal H}. 
\label{n62}
\end{equation}

\subsection{Liouville operator}
\label{s2.2}

The next task is to find out how the dynamics generated by the Hamiltonian $H_\lambda$, \eqref{n1}, acting on ${\mathbb C}^N$,  is represented in the Hilbert space ${\mathcal H}$. The answer is that 
\begin{equation}
\pi\big( e^{\i t H_\lambda} {\mathcal O}e^{-\i t H_\lambda}\big) = e^{\i t L_\lambda}\pi({\mathcal O})e^{-\i t L_\lambda},
\label{n63}
\end{equation}
for a self-adjoint operator 
\begin{equation}
L_\lambda = L_0 +\lambda I
\end{equation}
acting on $\mathcal H$, called the {\em Liouville operator}. It is the generator of dynamics in the purification Hilbert space,  akin to the Hamiltonian $H_\lambda$ without purification.  Given $H_\lambda$, the operator $L_\lambda$ satisfying \eqref{n63} is not unique. This is a general fact explained by the Tomita-Takesaki theory of von Neumann algebras \cite{BR}, which tells that the image of the map $\pi$ does not constitute all operators on ${\mathcal H}$ but only `half' of them in the following sense. There is an anti-unitary map $J$ on $\mathcal H$ (called the modular conjugation)  having the following property. Given any $A$ in the image of $\pi$ ({\em i.e.}, $A=\pi(a)$ for some operator $a$ on ${\mathbb C}^N\otimes {\mathcal F}$) the operator $JAJ$ commutes with all  operators in the image of $\pi$.\footnote{More precisely, if we denote by ${\mathfrak M}$ the weak closure of the image of $\pi$ and by ${\mathfrak M}'$ its commutant, then $J{\mathfrak M}J={\mathfrak M}'$.} We say that $JAJ$ is in the commutant of the algebra of observables. Now suppose a specific operator $L_\lambda$ satisfies \eqref{n63}. Then take any $X$ in the commutant and set  $K = L_\lambda+X$. This is again an operator on ${\mathcal H}$ and it also satisfies\footnote{This can be checked for instance by using the Kato-Trotter formula \cite{Zagrebnov}  -- note however that $L_\lambda$ is not in general an operator in the image of $\pi$, so $X$ does not commute with $L_\lambda$.} $\pi( e^{\i t H_\lambda} {\mathcal O}e^{-\i t H_\lambda}) = e^{\i t K}\pi({\mathcal O})e^{-\i t K}$. The non-uniqueness of the Liouville operator can be used to one's advantage by choosing a suitable operator $X$. 

Using the perturbation theory of equilibrium (KMS) states \cite{DJP,BR} it is known that the interacting system-reservoir complex has a unique equilibrium state, represented by a normalized vector $\Omega_{\s\r,\beta,\lambda}\in{\mathcal H}$. This vector has an explicit Taylor expansion in $\lambda$ at $\lambda=0$, and 
\begin{equation}
\Omega_{\s\r,\beta,\lambda} =\Omega_{\s,\beta}\otimes\Omega_\r + O(\beta\lambda),
\label{n66.1}
\end{equation}
where $\Omega_{\s,\beta}$ is given in \eqref{n47.1}. In other words, $\Omega_{\s\r,\beta,0}=\Omega_{\s,\beta}\otimes\Omega_\r$ is the non-interacting equilibrium state. We then use the above freedom (choice of $X$) to impose the condition
\begin{equation}
L_\lambda \Omega_{\s\r,\beta,\lambda}=0.
\label{n66.2}
\end{equation}
The resulting Liouville operator has the form \cite{JP,BFS,DJP,MAOP}
\begin{equation}
L_\lambda = L_0+\lambda I
\label{n66}
\end{equation}
where
\begin{eqnarray}
L_0 &=& L_\s +L_\r \label{3}\\
L_\s &=& H_\s\otimes\bbbone_\s\ - \bbbone_\s\otimes H_\s\\
L_\r &=&  d\Gamma(u) 
\label{4}
\end{eqnarray}
The system part $L_\s$ acts on ${\mathbb C}^N\otimes{\mathbb C}^N$ and $L_\r$ is the second quantization of multiplication by the radial variable $u\in{\mathbb R}$, acting on the Fock space ${\mathcal H}_\r$. In particular, $\e^{\i t L_\r} a_\beta(f)e^{-\i t L_\r} = a_\beta(e^{\i t u}f)$ (Bogoliubov transformation). The interaction operator $I$ in \eqref{n66} is given by $I = \pi\big(G\otimes \varphi(g)\big) - J\pi\big(G\otimes \varphi(g)\big)J$ (compare with \eqref{n1}), where $J$ is the modular conjugation. The explicit form of $J$ is well known, see {\em e.g.} (1.21) of \cite{KoMeSo}. One then gets the expression  \cite{KoMeSo,MAOP}
\begin{equation}
I = G\otimes \bbbone_\s\otimes \varphi_\beta(\tau_\beta g) - \bbbone_\s\otimes {\mathcal C} G{\mathcal C}\otimes \varphi_\beta(e^{-\beta u/2} \tau_\beta g).
\label{n40}
\end{equation}

\subsection{Resonance expansion and proof of Theorem \ref{thm1}}
\label{s2.3}

Let ${\mathcal O}$ be a system-reservoir observable. According to \eqref{n60}, \eqref{n62} and \eqref{n63}, its average at time $t$ in the state $\rho_\s\otimes\omega_{\r,\beta}$ is given by
\begin{equation}
\rho_\s\otimes\omega_{\r,\beta}({\mathcal O}) = \langle\Psi_\s\otimes\Omega_{\r}, e^{\i t L_\lambda}\pi({\mathcal O})e^{-\i tL_\lambda} (\Psi_\s\otimes\Omega_\r)\rangle.
\label{n71}
\end{equation}
Given any vector $\Psi_\s\in{\mathbb C}^N\otimes{\mathbb C}^N$, there exists a unique operator $B$ on ${\mathbb C}^N$ such that 
\begin{equation}
\Psi_\s = (\bbbone_\s\otimes B)\Omega_{\s,\beta} \in {\mathbb C}^N\otimes{\mathbb C}^N,
\label{n72}
\end{equation}
where $\Omega_{\s,\beta}$ is the purification vector \eqref{n47.1} representing the system Gibbs equilibrium state. The operator $B$ can be found explicitly and easily as follows (see also \cite{MAOP}). Since the density matrix $\rho_{\s,\beta}=Z^{-1}_{\s,\beta}e^{-\beta H_\s}$ is invertible, we can write (see also the beginning of Section \ref{s2.1}) $\rho_\s = \rho_{\s,\beta}\rho_{\s,\beta}^{-1}\rho_\s = \sum_jp_j|\psi_j\rangle\langle\psi_j| \rho_{\s,\beta}^{-1}\rho_\s = \sum_jp_j|\psi_j\rangle\langle\rho_\s \rho_{\s,\beta}^{-1}\psi_j|$. The latter sum has the representation $\sum_j\sqrt{p_j} \psi_j\otimes {\mathcal C}\rho_\s \rho_{\s,\beta}^{-1}\psi_j = (\bbbone_\s\otimes {\mathcal C}\rho_\s \rho_{\s,\beta}^{-1}{\mathcal C})\Omega_{\s,\beta}$ in the purification space ${\mathbb C}^N\otimes{\mathbb C}^N$. Hence $B={\mathcal C}\rho_\s \rho_{\s,\beta}^{-1}{\mathcal C}$. Note that 
\begin{equation}
\| B\| =\|\rho_\s \rho^{-1}_{\s,\beta}\| \le {\rm tr}_{{\mathbb C}^N} e^{\beta (\|H_\s\|-H_\s)}
\label{n43.1}
\end{equation}
since the operator norm $\|\rho_\s\|$ equals the biggest eigenvalue of $\rho_\s$ which does not exceed $1$ and  $\|\rho^{-1}_{\s,\beta}\|=Z_{\s,\beta} \max_{1\le j\le N} e^{\beta E_j}\le Z_{\s,\beta} e^{\beta \|H_\s\|} = e^{\beta \|H_\s\|}\, {\rm tr}_{{\mathbb C}^N} e^{-\beta H_\s} $ (see \eqref{n47.1}).

The point is that the operator
\begin{equation}
B'=\bbbone_\s\otimes B\otimes\bbbone_\r
\label{n73}
\end{equation}
commutes with the operator $e^{\i t L_\lambda}\pi({\mathcal O})e^{-\i tL_\lambda}$ (as the latter belongs to (the weak closure) of the range of $\pi$ and $B'$ belongs to the commutant of this range). We thus obtain from \eqref{n71}-\eqref{n73}
\begin{equation}
\rho_\s\otimes\omega_{\r,\beta}({\mathcal O}) = \langle\Psi_\s\otimes\Omega_{\r}, B' e^{\i t L_\lambda}\pi({\mathcal O})e^{-\i tL_\lambda} (\Omega_{\s,\beta}\otimes\Omega_\r)\rangle.
\label{n74}
\end{equation}
Next, using the normalization \eqref{n66.2}, the bound \eqref{n43.1} and the expansion \eqref{n66.1} yields
\begin{equation}
\rho_\s\otimes\omega_{\r,\beta}({\mathcal O}) = \langle\Psi_\s\otimes\Omega_{\r}, B' e^{\i t L_\lambda}\pi({\mathcal O}) (\Omega_{\s,\beta}\otimes\Omega_\r)\rangle + R_1,
\label{n76}
\end{equation}
where 
\begin{equation}
|R_1|\le C |\lambda| \|{\mathcal O}\|
\label{n77}
\end{equation}
is a constant independent of $\lambda, t$, $\rho_\s$ and $\mathcal O$ (we also used $\|e^{\pm\i t L_\lambda}\|=1$ and $\|\pi({\mathcal O})\|=\|{\mathcal O}\|$). Note that in \eqref{n76} we cannot commute $B'$ back to the right as it will not commute with $e^{\i t L_\lambda}\pi({\mathcal O})$, because that operator is not in the range of $\pi$ any longer, after cancelling the factor $e^{-\i tL_\lambda}$. 

Consider now an observable ${\mathcal O}=A\otimes\bbbone_{\mathcal F}$ on the system alone ($A$ an $N\times N$ matrix). Then the main term on the right side of \eqref{n76} is of the form as indicated in Corollary 3.2 
of \cite{Markov1},
\begin{eqnarray}
\langle\Psi_\s\otimes\Omega_{\r}, B' e^{\i t L_\lambda}\pi(A\otimes\bbbone_{\mathcal F}) (\Omega_{\s,\beta}\otimes\Omega_\r)\rangle &=& \langle\phi_\s\otimes\Omega_{\r}, e^{\i t L_\lambda} \psi_\s\otimes\Omega_\r\rangle \label{nn78}\\
\phi_\s &=& (\bbbone_\s\otimes B^*)\Psi_\s\\
\psi_\s &=& (A\otimes\bbbone_\s)\Omega_{\s,\beta}.\label{nn80}
\end{eqnarray}
Assuming that $0<|\lambda|^{3/4}\le c\lambda_0$ ({\em c.f.} (3.14) 
in \cite{Markov1}), the Corollary 3.2
of \cite{Markov1} thus yields
\begin{equation}
\Big| \langle\phi_\s \otimes\Omega_\r, e^{\i tL_\lambda}\psi_\s\otimes\Omega_\r \rangle_{\mathcal H}  - \langle\phi_\s, e^{\i t M(\lambda)}\psi_\s\rangle_{{\mathbb C}^N\otimes{\mathbb C}^N}\Big| \leq C |\lambda|^{1/4}\varkappa_0\, \|\phi_\s\|\, \|\psi_\s\|,
\label{n81}
\end{equation}
where
\begin{equation}
M(\lambda) = \bigoplus_{e\in{\mathcal E}_0} M_e,\qquad M_e = \sum_{s\in{\mathcal S}_e^{\rm dec}} \big(e+\lambda^2\aes\big) \Qes + \sum_{s\in{\mathcal S}^{\rm osc}_e} \Ees \Qes
\label{n82.1}
\end{equation}
and 
\begin{eqnarray}
\varkappa_0&=&
\max\Big[1, 1/g,\kappa/a,\alpha\kappa,\varkappa_1(1+\varkappa_1)(1+g+1/g),\nonumber\\
&&\varkappa_1(1+\varkappa_1^4)\kappa \max\big\{ 1,\tfrac{1+\alpha}{\delta}, 1/a, \tfrac{1+\kappa}{a\delta}, \kappa^2\big( \tfrac{\kappa}{a}(1+\kappa^3/\delta^3)(1+1/a)+1/\delta^2
\big)\big\}\Big]. \qquad
\label{160.1.1}
\end{eqnarray}
The various parameters appearing in \eqref{160.1.1} are defined in \eqref{FGR}-\eqref{g}.

Due to the Fermi Golden Rule condition, all ${\rm Im}\, a_e^{(s)}>0$ and $|{\mathcal S}_e^{\rm osc}|=1$. There is exactly one real eigenvalue of $L_\lambda$, namely $E_0^{(1)}(\lambda)=0$. This is also the only real eigenvalue of $\Lambda_0$, with associated eigenprojection   $Q_0^{(1)}=|\Omega_{\s,\beta}\rangle \langle\Omega_{\s,\beta}|$. It follows from \eqref{n82.1} that 
\begin{equation}
M(\lambda) = L_\s +\lambda^2 \bigoplus_{e\in{\mathcal E}_0}   \Lambda_e \equiv L_\s +\lambda^2 \Lambda.
\label{n83.1}
\end{equation}
Now
\begin{equation}
e^{\i tM(\lambda)}\psi_\s = e^{\i tM(\lambda)}(A\otimes\bbbone_\s)\Omega_{\s,\beta} = (A(t)\otimes\bbbone_\s)\Omega_{\s,\beta}
\label{n82}
\end{equation}
which defines the operator $A(t)$ uniquely. This is a consequence of the fact that any vector $\chi_\s\in{\mathbb C}^N\otimes{\mathbb C}^N$ can be written as $\chi_\s = (D\otimes\bbbone_\s)\Omega_{\s,\beta}$ for a unique operator $D$ on ${\mathbb C}^N$ (one says that the vector $\Omega_{\s,\beta}$ is cyclic for the algebra ${\mathcal B}({\mathbb C}^N)\otimes\bbbone_\s$). It is clear from \eqref{n82} that ${\mathbb R}_+\ni t\mapsto A(t)$ is a group and $A(0)=A$. Hence 
\begin{equation}
A(t)=e^{t{\mathcal L}_*}A,
\label{n83}
\end{equation} 
for a generator ${\mathcal L}_*$ acting on system observables  $A\in {\mathcal B}({\mathbb C}^N)$, the operators on ${\mathbb C}^N$. We  have ({\em c.f.} \eqref{nn78}-\eqref{nn80})
\begin{eqnarray}
\langle\Psi_\s\otimes\Omega_{\r}, B' e^{\i t L_\lambda}\pi(A\otimes\bbbone_\r) (\Omega_{\s,\beta}\otimes\Omega_\r)\rangle &=& \langle\Psi_\s, (\bbbone_\s\otimes B) (e^{t {\mathcal L}_*}A\otimes\bbbone_\s) \Omega_{\s,\beta}\rangle +R_2 \nonumber\\
&=&\langle\Psi_\s, (e^{t{\mathcal L}_*} A\otimes\bbbone_\s)\Psi_\s\rangle+R_2\nonumber\\
&=& {\rm tr}_{{\mathbb C}^N} \big(\rho_\s\,  (e^{t{\mathcal L}_*}A)\big) +R_2\nonumber\\
&=& {\rm tr}_{{\mathbb C}^N}\big(( e^{t{\mathcal L}} \rho_\s)\, A\big)+R_2
\label{n89}
\end{eqnarray}
where ${\mathcal L}$ is the adjoint of ${\mathcal L}_*$ \footnote{This is the adjoint with respect to the Hilbert space of Hilbert-Schmidt operators acting on ${\mathbb C}^N$, with inner product $\langle a,b\rangle = {\rm tr}_{{\mathbb C}^N}(a^*b)$. }
and 
\begin{equation}
|R_2|\le 	C|\lambda|^{1/4} \varkappa_0 \| (\bbbone_\s\otimes B^*)\Psi_\s\|\, \|(A\otimes\bbbone_\s)\Omega_{\s,\beta} \| \le C|\lambda|^{1/4} \|A\|,
\label{62}
\end{equation}
for a constant $C$ independent of $\rho_\s$ and independent of $\lambda, g, a, \alpha, \kappa, \delta$.  In the chain of equalities \eqref{n89} we have first used that $\bbbone_\s\otimes B$ commutes with $(e^{t {\mathcal L}_*}A)\otimes\bbbone_\s$, then that $(\bbbone_\s\otimes B)\Omega_{\s,\beta}=\Psi_\s$ (see \eqref{n72}) and finally that $\Psi_\s$ is the purification of the initial system density matrix $\rho_\s$ (see \eqref{n44.1}). 

We now combine \eqref{n89} with \eqref{n76}:
\begin{equation}
\rho_\s\otimes \omega_{\r,\beta}(A\otimes\bbbone_{\mathcal F}) = {\rm tr}_{{\mathbb C}^N} \big((e^{t{\mathcal L}}\rho_\s) A\big)
+R_1+R_2.
\label{63.1}
\end{equation}
The bounds \eqref{n77} and \eqref{62} yield
\begin{equation}
|R_1+R_2| \le C|\lambda|^{1/4} \|A\|.	
\label{64}
\end{equation}
According to \eqref{10}, \eqref{11} we have  $\rho_\s\otimes\omega_{\r,\beta}(A\otimes\bbbone_{\mathcal F}) = {\rm tr}_{{\mathbb C}^N} ((V_t\rho_\s)A)$ and so \eqref{63.1}, \eqref{64} give 
\begin{equation}
\Big| {\rm tr}_{{\mathbb C}^N} \big( (V_t\rho_\s)A- (e^{t\mathcal L}\rho_\s)A\big)\Big|\le C|\lambda|^{1/4} \|A\|.
\end{equation}
It follows that 
\begin{equation}
\big\| V_t\rho_\s - e^{t{\mathcal L}}\rho_\s \big\|\le C|\lambda|^{1/4},
\label{65}
\end{equation}
where the norm is that of functionals on ${\mathcal B}({\mathbb C}^N)$. Upon changing the constant $C$ in \eqref{65}, the norm can also be taken to be the trace norm (equivalence of norms in finite dimensional spaces). Since $C$ does not depend on $\rho_\s$, \eqref{65} gives $\| V_t-e^{t{\mathcal L}}\|\le C|\lambda|^{1/4}$. By definition, ${\mathcal L}$ is  the adjoint of ${\mathcal L}_*$, \eqref{n83}, and it is shown in the Appendix of \cite{MAOP} that ${\mathcal L}=-\i [H_\s,\cdot]+\lambda^2K$, where $K$ is the Davies generator.   This completes the proof of Theorem \ref{thm1}.\qed

\section{Verifying conditions (A1)-(A5) of \cite{Markov1}}
\label{versect}

The Hilbert space in the current paper has the form ${\mathcal H}={\mathcal H}_\s\otimes{\mathcal H}_\r$ and $\dim{\mathcal H}_\s<\infty$. The uncoupled operator is $L_0 = L_\s\otimes\bbbone_\r + \bbbone_\s\otimes L_\r$, see \eqref{3}. We will often leave out the trivial factors $\bbbone_\s$, $\bbbone_\r$ in our notation. The assumption (A1) of \cite{Markov1} is that all eigenvalues of $L_\lambda$ for small $\lambda\neq 0$ are simple. This is true in the current setup, since the only eigenvalue of $L_\lambda$ for $\lambda\neq 0$ small is $e=0$ and it is simple, as is guaranteed by the Fermi Golden Rule Condition \eqref{n97} (see also the Remark (ii) following it). The condition (A4) of \cite{Markov1} is that $P_eIP_e=0$, which is obviously true.

\subsection{ Regularity of the resolvent, conditions (A2), (A3)}

The reservoir dynamics has a unique stationary state represented by $\Omega_\r\in{\mathcal H}_\r$, that is, ${\rm Ker} L_\r={\mathbb C}\Omega_\r$. This is clear from \eqref{4}. We set 
\begin{equation}
P_\r=\bbbone_\s\otimes|\Omega_\r\rangle\langle\Omega_\r| \qquad \mbox{and} \qquad P^\perp_\r=\bbbone_{\mathcal H}- P_\r.
\label{n58}
\end{equation}
We denote by $\bar {\mathcal O}$ is the restriction of an operator $\mathcal O$ to the range of $P_\r^\perp$: $\bar {\mathcal O}=P^\perp_\r {\mathcal O}P^\perp_\r \upharpoonright_{{\rm Ran} P^\perp_\r}$. For $z\in\mathbb C\backslash \mathbb R$ the reduced resolvent is given by
$$
R_z^{P_\r}(\lambda) = (P^\perp_\r L_\lambda P^\perp_\r-z)^{-1}\upharpoonright_{{\rm Ran}P^\perp_\r} \equiv (\bar L_\lambda -z)^{-1}.
$$  
For any integer $j\ge 0$, set
\begin{equation}
\label{clgamma}
\gamma_j(\beta)\equiv \gamma_j=  \|\partial^j_u g_\beta\|_{L^2({\mathbb R}\times S^2)},
\end{equation}
where we set for short (see \eqref{n50})
\begin{equation}
g_\beta\equiv	g_\beta(u,\Sigma)\equiv (\tau_\beta g)(u,\Sigma)\in L^2({\mathbb R}\times S^2)\equiv L^2({\mathbb R}\times S^2, du\times d\Sigma).
\label{61}
\end{equation}

We note that smoothness of $g_\beta$ implies decay of the reservoir correlation function. More precisely, the (symmetrized) reservoir correlation function \eqref{16.2} is given by 
\begin{equation}
c_\beta(t) = \int_{\mathbb R} du\  e^{\i t u}\int_{S^2} d\Sigma \ |g_\beta(u,\Sigma)|^2, 
\label{98}
\end{equation}
see {\em e.g.} \cite{MAOP}, Discussion point (2) after (2.17). Then, if $\gamma_j$ is finite for $j=0,\ldots,k$, we can use  $e^{\i tu}= (\i t)^{-1}\partial_ue^{\i tu}$ and integrate by parts in \eqref{98} $k$ times. This shows that $c_\beta(t)$ decays at least as $t^{-k}$. As we show below in \eqref{93}, we need $k=4$ for our method, which implies the decay behaviour \eqref{18}.

Define the norm  
\begin{equation}
\|\phi\|_j= \|(1+\bar A^2)^{j/2}\phi\|,
\end{equation} 
on $\mathcal H$, where 
\begin{equation}
A=d\Gamma(\i\partial_u).
\label{nA}
\end{equation}
Note that $\|\phi\|_k\le\|\phi\|_\ell$ for $0\le k\le \ell$. Denote by ${\mathbb C}_- =\{ z\in {\mathbb C} : {\rm Im} z<0\}$ the open lower complex half plane.  

\begin{thm}
	\label{thm0}
	Take $\lambda$ and $\beta$ in a range such that $|\lambda|\, \|\partial^2_ug_\beta\|\le c$ for some constant $c>0$ independent of $\lambda, \beta$. Then we have for all integers $k\ge 0$:
	\begin{eqnarray}
	\sup_{z\in{\mathbb C}_-} \big| \partial^k_z \langle\phi, R_z^{P_\r}(\lambda)\psi\rangle\big|  &\le& C\big(1+|\lambda|e^{|\lambda|\gamma_2}(\gamma_2+\gamma_3+\cdots+\gamma_{k+2})\big) e^{|\lambda|\gamma_2} \|\phi\|_{k+1} \|\psi\|_{k+1},\qquad 
	\label{031.1.1}\\
	\sup_{z\in{\mathbb C}_-} \big| \partial_\lambda \langle\phi, R_z^{P_\r}(\lambda)\psi\rangle\big|&\le& C\big( \gamma_0 +(\gamma_2+\gamma_0\gamma_3) e^{|\lambda|\gamma_2} \big)e^{|\lambda|\gamma_2}  \|\phi\|_2\|\psi\|_2,
	\label{031.1}
	\end{eqnarray}
	where $C$ is a constant independent of $\lambda,\beta$. 
\end{thm}
Theorem \ref{thm0} shows that condition (A2) of \cite{Markov1} holds provided 
\begin{equation}
\gamma_j = \|\partial^j_u g_\beta\|_{L^2({\mathbb R}\times S^2)}<\infty, \qquad \mbox{for $j=0,1,2,3,4$}.
\label{93}
\end{equation} 
In particular, if \eqref{93} holds, then we have
\begin{eqnarray}
\max_{0\le j\le 2}\sup_{z\in{\mathbb C}_-}\Big| \partial_z^j \langle\phi, R_z^{P_\r}(\lambda) \psi\rangle\Big| &\le& C_1(\phi,\psi)<\infty,
\label{m03}\\
\sup_{z\in{\mathbb C}_-}\Big| \partial_\lambda\langle\phi, R_z^{P_\r}(\lambda) \psi\rangle\Big| &\le& C_1(\phi,\psi)<\infty,
\label{m03.1}
\end{eqnarray}
for all vectors $\phi$, $\psi$ in the dense set 
\begin{equation}
{\mathcal D} = \{\phi\in{\mathcal H}\ :\ \|\phi\|_3<\infty \}.
\label{n67.1}
\end{equation}
Furthermore, it follows from \eqref{031.1.1} and \eqref{031.1} that 
\begin{eqnarray}
C_1(\phi,\psi) &\prec& c_1(\lambda,\beta)\|\phi\|_3\|\psi\|_3\nonumber\\
c_1(\beta,\lambda)&=& e^{|\lambda|\gamma_2} \Big[ 1+\gamma_0 +e^{|\lambda|\gamma_2}\big(\gamma_2+\gamma_0\gamma_3 +|\lambda|(\gamma_2+\gamma_3+\gamma_4)\big)\Big].
\label{n67}
\end{eqnarray}
The symbol $A\prec B$ means that $A\le {\rm const.} B$ for a constant const. independent of $\lambda,\beta$. 

\smallskip

Using the explicit form of $I$, \eqref{n40}, we get that for any vector $\phi\in{\rm Ran}P_\r$,
\begin{eqnarray}
\|I P_\r \phi \|_j  &\le& \sqrt 2 \|G\| \, \|(1+A^2)^{j/2} a^*(g_\beta)\Omega_\r\|\, \|\phi\|
=\sqrt 2 \|G\| \,  \| (1-\partial_u^2)^{j/2} g_\beta\|\,  \|\phi\|\nonumber\\
&\prec&  \big(\gamma_0+\gamma_j \big) \|\phi\|.
\label{88}
\end{eqnarray}
Consequently, we have (see the definition (3.12)
of \cite{Markov1})
\begin{equation}
\varkappa_1 = \max_{m,n} C_1\big(I(\varphi_m\otimes\Omega_\r), I(\varphi_n\otimes\Omega_\r)\big)\prec c_1(\beta,\lambda) (\gamma_0+\gamma_3)^2.
\label{89}
\end{equation}

The condition (A3) of  \cite{Markov1} is that $IP_e$ is bounded and belongs to the dense set $\mathcal D$, \eqref{n67.1}, which is immediately seen to hold provided \eqref{93} holds with $j=0,1,2,3$ only.  
\smallskip

We now show how the Assumptions \eqref{gcond}, \eqref{paramvalues} imply \eqref{93}. Recall the definition \eqref{n50} of $g_\beta=\tau_\beta g$. Note that the square root $\sqrt{u/(1-e^{-\beta u})}$ is analytic at $u=0$. It is then clear that if $g$ satisfies \eqref{gcond} with $p=-\tfrac12, \tfrac12, \tfrac32$ then $\gamma_j$ is finite for $j=0,1,2,3,4$. The point of the discrete powers of $p$ is that they combine with $|u|^{1/2}$ in \eqref{n50} to yield a smooth function at $u=0$. For vaues $p>2$, we use Lemma \ref{lemma1} below to see that $\gamma_j<\infty$ for $j=0,1,2,3,4$. This lemma gives also a bound on $\gamma_j$ with an explicit temperature dependence. 
\begin{lem}
	\label{lemma1}
	Denote the $\ell$th radial derivative $\partial^\ell_{|k|}g(k)$ simply by $g^{(\ell)}(k)$, $k\in{\mathbb R}^3$, with the convention that $g^{(-1)}\equiv 0$. We have for all integers $j\ge 0$
	\begin{eqnarray}
	\gamma_j  &\le& C \max_{0\le r\le j}\Big(\beta^r \big\{ \big\| g^{(j-r)}\big\| +  \big\| |k|^{-1} g^{(j-r-1)}\big\| \big\}\nonumber\\
	&& \qquad\qquad  +\beta^{-1/2}\big\{   \big\| |k|^{-r-1/2}g^{(j-r)}\big\| +  \big\| |k|^{-r-3/2}g^{(j-r-1)}\big\|\big\}\Big),\quad 
	\label{thebound}
	\end{eqnarray}
	where $C$ is independent of $\beta$ and  the norm on the right side is that of $L^2({\mathbb R}^3,d^3k)$. 
\end{lem}

To satisfy condition (A3) of \cite{Markov1} it suffices to impose that the right side of \eqref{thebound} be bounded for $j=3$. This covers the case $p>2$ in \eqref{gcond}, \eqref{paramvalues}.
\medskip

\noindent
{\em Proof of Lemma \ref{lemma1}. }  Set  
\begin{equation}
D_0=|1-e^{-\beta u}|^{-1/2},\quad D_r = \partial_u^r D_0,\quad \gamma(u,\Sigma) =  |u| \left\{
\begin{array}{ll}
g(u,\Sigma), & u\ge 0\\
-\bar g(-u,\Sigma), & u<0
\end{array}
\right. .
\label{125}
\end{equation}
For $u\neq 0$, we have $D_1=\frac{\beta}{2} D_0[1-{\rm sgn}(u)D^2_0]$, where ${\rm sgn}(u)=u/|u|$ is the sign of $u$. Then $D_r$ is a polynomial of degree $2r+1$ in $D_0$ and so $|D_r|\le C\beta^r(|D_0| + |D_0|^{2r+1})$ for a $C$ independent of $\beta$ and $u$. Now $|D_0| \le 1+(\beta |u|)^{-1/2}$ and hence $|D_r|\le C\beta^r(1+ (\beta|u|)^{-r-1/2})\le C(\beta^r +\beta^{-1/2} |u|^{-r-1/2})$. Then 
\begin{equation}
\|\partial_u^j g_\beta\| \le C \sum_{r=0}^j \|D_r\partial_u^{j-r}\gamma\| \le C \max_{0\le r\le j}\Big(\beta^r \|\partial^{j-r}_u\gamma\| + \beta^{-1/2} \big\| |u|^{-r-1/2}\partial_u^{j-r}\gamma\big\| \Big).
\label{126}
\end{equation}
Next, from \eqref{125}, $|\partial_u^\ell\gamma|\le C \big( |u| |(\partial^\ell_u g){(|u|,\Sigma)} |+ |(\partial_u^{\ell-1}g)(|u|,\Sigma)|\big)$ (with $u\neq 0$). Using this bound, we have in terms of Euclidean coordinates $k\in{\mathbb R}^3$, 
\begin{eqnarray}
\big\| |u|^s \partial_u^\ell\gamma\big\|^2 &\le& C \int_{-\infty}^\infty du\int_{S^2}d\Sigma \ |u|^{2s}\big( u^2 |(\partial^\ell_u g){(|u|,\Sigma)}|^2 + |(\partial_u^{\ell-1}g)(|u|,\Sigma)|^2\big) \nonumber\\
&=& 2C\int_0^\infty du \int_{S^2}d\Sigma \ u^{2s}\big( u^2 |(\partial^\ell_u g){(u,\Sigma)}|^2 + |(\partial_u^{\ell-1}g)(u,\Sigma)|^2\big) \nonumber\\
&=& 2C\int_{{\mathbb R}^3} d^3k  \ |k|^{2s}\big(| g^{(\ell)}(k)|^2 + |k|^{-2} |g^{(\ell-1)}(k)|^2\big) \nonumber\\
&=& 2 C \Big( \big\| |k|^s g^{(\ell)}\big\|^2+ \big\| |k|^{s-1} g^{(\ell-1)}\big\|^2\Big). 
\label{127.1}
\end{eqnarray}
For $\ell=0$, the estimate holds upon setting $g^{(-1)}\equiv 0$. 
Combining \eqref{126} and \eqref{127.1} gives \eqref{thebound}. This completes the proof of Lemma \ref{lemma1}. \qed

\subsubsection{Proof of Theorem \ref{thm0}}

The proof follows the strategy of \cite{KM1,KoMeSo}, however, in those papers, the interaction operator $I$ is proportional to a Weyl operator (which is, in particular, a bounded operator) and not the field operator we have here. It is thus necessary to give a proof in the present case. Introduce the regularized Liouville operator $\bar L(\alpha)$, for $\alpha>0$, by
\begin{equation}
\label{t1}
\bar L(\alpha) = \bar L_0 +\i \alpha \bar N +\lambda\bar I(\alpha),
\end{equation}
where $\bar X$ means the restricted operator $P^\perp_\r XP^\perp_\r\upharpoonright_{{\rm Ran}P^\perp_\r}$, and where 
\begin{equation}
\label{t2}
\bar I(\alpha) = \frac{1}{\sqrt{2\pi}} \int_{\mathbb R} \widehat f(s) e^{\i s\alpha \bar A} \bar I e^{-\i s\alpha \bar A}ds , \qquad A=d\Gamma(\i\partial_u).
\end{equation}
Here, $\widehat f(s)=(2\pi)^{-1/2}\int_{\mathbb R} e^{-\i sx}f(x)dx$ is the Fourier transform of the function $f$, which we choose to be a Schwartz function satisfying $f^{(k)}(0)=1$ for $k=0,1,\ldots$ This implies
\begin{equation}
\frac{1}{\sqrt{2\pi}} \int_{\mathbb R} (\i s)^k \widehat f(s) ds =1,\qquad k=0,1,\ldots
\label{schwartz}
\end{equation}	
In what follows, we use the notation
\begin{equation}
\label{t5}
R_z(\alpha) = (\bar L(\alpha)-z)^{-1}.
\end{equation}

\begin{prop}
	\label{prop1}
	Suppose  $8|\lambda| ( \int_{{\mathbb R}}|s \widehat f(s) |ds)\|G\|\, \|\partial_ug_\beta\| \le \sqrt{2\pi}$. Then the following holds:
	\begin{itemize}
		\item[{\em 1.}] We have  $\frac{\alpha}{2} \bar N\le {\rm Im}\bar L(\alpha) \le 2\alpha\bar N$. In particular, $z$ with ${\rm Im} z <\alpha/2$ is in the resolvent set of $\bar L(\alpha)$ and $\|R_z(\alpha)\|\le (\alpha/2-{\rm Im} z)^{-1}$. 
		
		\item[{\em 2.}] We have $\|\bar N^{1/2} R_z(\alpha)\psi\|\le \sqrt{\frac2\alpha} |\langle \psi, R_z(\alpha)\psi\rangle|^{1/2}$ and $\|\bar N^{1/2} R_z(\alpha)\psi\|\le \frac{2}{\alpha}\|\bar N^{-1/2}\psi\|$ as well as $\|\bar N^{1/2} R_z(\alpha)\bar N^{1/2}\|\le\frac{2}{\alpha}$. The same bounds hold for $R_z(\alpha)$ replaced by $R_z(\alpha)^*$. 
		
		\item[{\em 3.}] For $z$ with ${\rm Im}z<0$ the range of  $R_z(\alpha)$ belongs to the domain of $\bar N$ and of $\bar L_0$ an moreover, $R_z(\alpha)$ leaves the domain of $\bar A$ invariant. Furthermore, $R_z(\alpha)$ converges to $(\bar L-z)^{-1}$ in the strong sense on ${\rm Ran}P_\r^\perp$ as $\alpha\rightarrow 0_+$. 
		
	\end{itemize}
\end{prop}

\bigskip

{\em Proof of Proposition \ref{prop1}.\ }  The arguments are similar to Lemma 4.3 of \cite{KoMeSo}, the difference being that there, the interaction operator is a Weyl operator as opposed to the field operator we have here. Once statement 1 is proven, the other statements follow from it as in \cite{KoMeSo}. We then only show statement 1 here. To do so, write
\begin{equation}
\label{t3}
{\rm Im}\bar L(\alpha) =\alpha \bar N^{1/2} \Big( \bbbone+\frac{\lambda}{\sqrt{2\pi}\, \alpha} {\rm Im}\int_{{\mathbb R}}\widehat f(s)\bar N^{-1/2}\big[ e^{\i s\alpha\bar A}\bar I e^{-\i s\alpha \bar A} -\bar I\, \big] \bar N^{-1/2} ds
\Big) \bar N^{1/2}.
\end{equation}
By writing $ e^{\i s\alpha\bar A}\bar I e^{-\i s\alpha \bar A} -\bar I = \int_0^{s\alpha}\partial_{s'}( e^{\i s'\bar A}\bar I e^{-\i s'\bar A}) ds'$ we get the estimate
\begin{equation}
\label{t4}
\big\| \bar N^{-1/2}\big[ e^{\i s\alpha\bar A}\bar I e^{-\i s\alpha \bar A} -\bar I\, \big] \bar N^{-1/2}\| \le \alpha|s| \|\bar N^{-1/2}[\bar A,\bar I]\bar N^{-1/2}\| \le 4\alpha|s| \|G\|\, \|\partial_u g_\beta\|_{L^2}.
\end{equation}
Combining \eqref{t3} and \eqref{t4} we get the bounds on ${\rm Im}\bar L(\alpha)$. The bound on the resolvent follows easily. This completes the proof of Proposition \ref{prop1}. \qed

\begin{prop}
	\label{prop2}
	Suppose that  for some $\ell\ge 0$, we have $\|\partial_u^kg_\beta\|<\infty$ for $k=1,\ldots,\ell+1$. 
	\begin{itemize}
		\item[{\em 1.}]
		Set $\partial I(\alpha)= \frac{d}{d\alpha}\bar I(\alpha) -[\bar A,\bar I(\alpha)]$. Then
		\begin{equation}
		\| \bar N^{-1/2} \partial I(\alpha)\| \le c_\ell \alpha^\ell, 
		\label{t8}
		\end{equation}
		where 
		\begin{equation}
		\label{cl}
		\ \  c_\ell= \frac{4 \|G\| }{\ell! \sqrt{ 2\pi}}  \|\partial^{\ell+1}_u g_\beta\|\,  \int_{{\mathbb R}} (1+|s|) \, \big| s^\ell\widehat f(s)\big| ds.
		\end{equation}

		\item[{\em 2.}] We have
		\begin{equation}
		\label{t9}
		\|\bar N^{1/2} R_z(\alpha)\psi\| \le\frac{2\sqrt 2}{\sqrt \alpha}  e^{4+|\lambda| c_1}\, \|\bar N^{-1/2}(1+\bar A^2)^{1/2}\psi\|, \quad c_1=c_{\ell=1}
		\end{equation}
	\end{itemize}
\end{prop}

\bigskip

{\em Proof of Proposition \ref{prop2}. }  1. We have
\begin{equation}
\partial I(\alpha) = \frac{1}{\sqrt{2\pi}}\int_{{\mathbb R}}(\i s-1)\widehat f(s) e^{\i s\alpha\bar A} [\bar A,\bar I] e^{-\i s\alpha\bar A}ds.
\label{t7}
\end{equation}
We denote the $l$-fold commutator ${\rm ad}_{\bar A}^l(\bar I)=[\bar A, [ \bar A, [,\cdots,[\bar A,\bar I]\cdots]]]$ and Taylor expand around $s=0$,
\begin{equation}
e^{\i s\alpha\bar A} [\bar A,\bar I] e^{-\i s\alpha\bar A} = \sum_{r=0}^\ell\frac{(\i s \alpha)^r}{r!}  \, {\rm ad}_{\bar A}^{r+1}(\bar I) +\frac{(\i s \alpha)^{\ell+1}}{(\ell+1)!} e^{\i s'\alpha\bar A} {\rm ad}_{\bar A}^{\ell+2}(\bar I) e^{-\i s'\alpha\bar A}, 
\label{t6}
\end{equation}
for some $s'\in (0,s)$. When inserted into \eqref{t7}, the part coming from the sum $r=0,\ldots,\ell$ in \eqref{t6} integrates to zero due to \eqref{schwartz}. Next, $\|\bar N^{-1/2} {\rm ad}_{\bar A}^{\ell+2}(\bar I)\| \le 4\|G\| \, \|\partial^{\ell+2}_ug_\beta\|$. The estimate \eqref{t8} follows. 

2. Following the proof of Lemma 4.5 in \cite{KoMeSo} and keeping track of the constants, one derives $|\langle \varphi, R_z(\alpha)\varphi\rangle| \le 4 e^{8+ 2|\lambda|c_1} \  \|\bar N^{-1/2} (1+\bar A^2)^{1/2}\varphi\|^2$. (The exponential comes from a Gronwall type estimate.) Combining this with Proposition \ref{prop1} (2), the bound \eqref{t9} follows. This completes the proof of Proposition \ref{prop2}. \qed
\bigskip

{\bf Proof of Theorem \ref{thm0}.} We show the bounds \eqref{031.1.1}, \eqref{031.1} for the regularized resolvent $R_z(\alpha)$. Since the upper bounds are independent of $\alpha$, since $R_z$ is bounded uniformly in $\alpha$ for $z\in{\mathbb C}_-$ fixed, and since $R_z(\alpha)\rightarrow R_z$ strongly as $\alpha\rightarrow 0$ (Proposition \ref{prop1} (3)), the result follows. We write simply $R=(\bar L(\alpha)-z)^{-1}$ here.  Let $k\ge 0$ be an integer. We have $\partial_z^kR = k! R^{k+1}$ and, denoting  
\begin{equation}
\partial \equiv \frac{d}{d\alpha} -[\bar A,\cdot]
\end{equation} 
(see also Proposition \ref{prop2}, point 1), we get
\begin{equation}
\frac{d}{d\alpha} \langle \phi, R^{k+1} \psi\rangle = \langle A\phi, R^{k+1} \psi\rangle -\langle (R^*)^{k+1}\phi, A\psi\rangle +\langle \phi, \partial(R^{k+1}) \psi\rangle. 
\label{117}
\end{equation}
All operators are on the range of $P_\r^\perp$ and we write simply $X$ for $\bar X$ for operators $X$ ({\em c.f.} after \eqref{t1}). The derivation $\partial$ satisfies the Leibniz rule, and we have $\partial R = -\lambda R (\partial I) R$, because $\partial(\bar L_0+\i\alpha\bar N)=0$. We get
\begin{equation}
\partial (R^{k+1}) = \sum_{j=0}^{k} R^j (\partial R) R^{k-j} = -\lambda \sum_{j=1}^{k+1} R^j (\partial I) R^{k-j+2}.
\end{equation}
We first use $\|R\|\le 2/\alpha$ and \eqref{t9} to estimate 
\begin{equation}
|\langle A\phi, R^{k+1}\psi\rangle|\le  \|A\phi\|\ \| R^{k+1}\psi\|\le C  e^{|\lambda|c_1} \alpha^{-k-1/2}  \|\phi\|_{1} \|\psi\|_{1}.
\label{123}
\end{equation}
The same bound holds for $|\langle (R^*)^k\phi, A\psi\rangle|$. Next, using \eqref{t8}, \eqref{t9}, $\| R\|\le 2/\alpha$ and  $\|\bar N^{1/2} R\|\le 2/\alpha$ (Proposition \ref{prop1} (2.))
\begin{eqnarray}
\Big| \lambda \sum_{j=1}^{k+1} \langle \phi, R^j (\partial I) R^{k-j+2}\psi\rangle\Big| &\le&	 |\lambda|  \sum_{j=1}^{k+1} \|\bar N^{1/2} R^{j-1}\| \|R^*\phi\| \|\bar N^{-1/2}\partial I\| \|R^{k-j+1}\| \|R\psi\|\nonumber\\
&\le& C|\lambda| e^{2|\lambda|c_1} c_\ell \, \alpha^{\ell -k-1} \|\phi\|_1\ \|\psi\|_1. 
\label{124}
\end{eqnarray}
Combining \eqref{117}, \eqref{123} and \eqref{124} with $\ell=1$ gives
\begin{equation}
\big| \frac{d}{d\alpha}\langle\phi, R^{k+1}\psi\rangle\big| \le  C\big(1+|\lambda|c_1 e^{|\lambda|c_1}\big) e^{|\lambda|c_1}  \alpha^{-k-1/2} \|\phi\|_{1} \|\psi\|_{1}.
\label{119.2}
\end{equation}
We integrate the inequality \eqref{119.2} and use that $\|R|_{\alpha=1}\|\le 2$ (Proposition \ref{prop1} (1.)) to obtain
\begin{equation}
\big| \langle\phi, R^{k+1}\psi\rangle\big| \le C\big(1+|\lambda|c_1 e^{|\lambda|c_1}\big) e^{|\lambda|c_1} (1+ \alpha^{-k+1/2}) \|\phi\|_{1} \|\psi\|_{1}.
\label{120.1}
\end{equation}
Now we use \eqref{120.1} to get $|\langle A\phi, R^{k+1}\psi\rangle|\le  C\big(1+|\lambda|c_1 e^{|\lambda|c_1}\big) e^{|\lambda|c_1} (1+ \alpha^{-k+1/2}) \|\phi\|_{2} \|\psi\|_{1}$ 
and similarly for the second term on the right side of \eqref{117}, which then amounts to (use now $\ell=2$ in \eqref{124})
\begin{equation}
\big| \frac{d}{d\alpha}\langle\phi, R^{k+1}\psi\rangle\big| \le C \big(1+|\lambda|e^{|\lambda|c_1}(c_1+c_2) \big) e^{|\lambda|c_1}  (1+\alpha^{-k+1/2}) \|\phi\|_{2} \|\psi\|_{2}.
\label{119.3}
\end{equation}
Upon integration, \eqref{119.3} yields the bound
\begin{equation}
\big| \langle\phi, R^{k+1}\psi\rangle\big| \le C\big(1+|\lambda|e^{|\lambda|c_1}(c_1+c_2)\big) e^{|\lambda|c_1} (1+ \alpha^{-k+3/2}) \|\phi\|_{2} \|\psi\|_{2},
\label{120.2}
\end{equation}
which is \eqref{120.1} with the power of $\alpha$ increased by one and the norm of the vectors (and the prefactor) changed. We now iterate this procedure $k$ times to obtain \eqref{031.1.1}. 

Now we show \eqref{031.1}.  Since $\frac{d}{d\lambda}R=-RIR$, we investigate
\begin{equation}
\frac{d}{d\alpha} \langle \phi, RIR\psi\rangle =  \langle A\phi, RIR\psi\rangle -  \langle (RIR)^*\phi, A\psi\rangle +  \langle \phi, \partial(RIR)\psi\rangle.
\label{129}
\end{equation}
Using \eqref{t9},
\begin{equation}
|\langle A\phi, RIR\psi\rangle|\le \|\bar N^{1/2}R^*A\phi\| \ \|\bar N^{-1/2} I\|\ \|R\psi\| \le C\|g_\beta\| e^{|\lambda|c_1} \alpha^{-1/2}\|\phi\|_{2}\|\psi\|_2
\label{111000}
\end{equation}
and similarly for the second term on the right side of \eqref{129}. Next, 
\begin{eqnarray}
\lefteqn{
	| \langle \phi, \partial(RIR)\psi\rangle|}\nonumber\\
&\le& |\lambda|\ | \langle \phi, R(\partial I)RIR\psi\rangle|+ | \langle \phi, R(\partial I)R\psi\rangle| + |\lambda|\  | \langle \phi, RIR(\partial I)R\psi\rangle|.
\label{131}
\end{eqnarray}
Now 
\begin{eqnarray}
| \langle \phi, R(\partial I)RIR\psi\rangle| &\le& \|\bar N^{1/2}R^*\phi\|\ \|\bar N^{-1/2} \partial I\| \ \|R\|\ \| I\bar N^{-1/2}\| \ \|\bar N^{1/2}R\psi\| \nonumber\\
&\le & C\|g_\beta\| e^{2|\lambda|c_1} c_\ell \alpha^{\ell -2} \|\phi\|_1\ \|\psi\|_1.
\label{132}
\end{eqnarray}
The last term on the right side of \eqref{131} has the same upper bound. Next,
\begin{equation}
| \langle \phi, R(\partial I)R\psi\rangle| \le \|\bar N^{1/2} R^*\phi\|\ \|\bar N^{-1/2}\partial I\|\ \|R\psi\|\le Ce^{2|\lambda|c_1} c_\ell \alpha^{\ell-1} \|\phi\|_{1}\ \|\psi\|_1.
\label{133}
\end{equation}
Taking $\ell=2$ in \eqref{132} and $\ell=1$ in \eqref{133} and using this in \eqref{131} yields 
\begin{equation}
| \langle \phi, \partial(RIR)\psi\rangle|\le C e^{2|\lambda|c_1} \big( c_1+c_2\|g_\beta\| \big) \|\phi\|_1\|\psi\|_1.
\label{134}
\end{equation}
Combining \eqref{129}, \eqref{111000}  and \eqref{134} shows that 
\begin{equation}
\big| \frac{d}{d\alpha} \langle \phi, RIR\psi\rangle \big| \le Ce^{|\lambda|c_1} \big(\|g_\beta\| +(c_1+c_2\|g_\beta\|) e^{|\lambda|c_1} \big) \alpha^{-1/2} \|\phi\|_2\|\psi\|_2.
\label{135}
\end{equation}
Now $|\langle\phi, RIR|_{\alpha=1}\psi\rangle|\le C\|g_\beta\| \|\phi\| \|\psi\|$ (Proposition \ref{prop1} (2.)) and so integrating \eqref{135} gives 
$$
| \langle \phi, RIR\psi\rangle | \le Ce^{|\lambda|c_1} \big( \|g_\beta\| +(c_1+c_2\|g_\beta\|) e^{|\lambda|c_1} \big) \|\phi\|_2\|\psi\|_2.
$$
Since $\frac{d}{d\lambda}R=-RIR$, this shows \eqref{031.1}. The proof of Theorem \ref{thm0} is complete. \qed

\subsection{Proof of Proposition \ref{prop1.2}.\ }

By the expression $A\prec B$ we mean that $A\le {\rm const.} B$, where the constant is independent of the parameters 
$\lambda, g, a, \alpha, \kappa, \delta$ and of the inverse temperature $\beta$. According to the proof of Theorem\ref{thm1}, \eqref{n81}, we need that $0<|\lambda|\prec \lambda_0^{4/3}$, where $\lambda_0$ is given by (see \cite{Markov1}, equation (3.14))
\begin{equation}
\lambda_0 = \frac{\min\Big[ 1,a,\delta/\kappa^2, \|I P_\r\|, g^{3/2}\Big]}{\max\Big[ 1,\varkappa_1\kappa (1+\varkappa_1\kappa/\delta), \alpha,\varkappa_1\Big]}.
\end{equation}
The various parameters are given by (see \cite{Markov1}, (3.8)-(3.11)) 
\begin{eqnarray}
a&=& \min_{e,s}\big\{ {\rm Im}\aes : \aes\not\in{\mathbb R} \big\}  \qquad \mbox{(smallest nonzero imaginary part)}\label{FGR}\\
\alpha&=&  \max_{e,s} |\aes| = \max_e {\rm spr}(\Lambda_e)\qquad \mbox{(maximal spectral radius)}\label{alpha}\\
\delta&=&	\min_{e, s,s'}\big\{|a_e^{(s)} -a_e^{(s')}| : s\neq s'\big\}  \label{delta} \qquad \mbox{(gap in spectrum of the  $\Lambda_e$)}\\
\kappa&=& 	\max_{e,s} \|\Qes\|
\label{kappa}\\
g &=& \min\{ |e-e'|\ :\ e,e'\in {\rm spec}L_\s, e\neq e'\}\label{g}
\end{eqnarray}
We now estimate $\lambda_0$ in terms of the inverse temperature $\beta$. From \eqref{thebound}, $\gamma_j\prec \beta^{-1/2} +\beta^j$. We estimate $\varkappa_1$ using \eqref{89}. Imposing the constraint
\begin{equation}
|\lambda|\prec \min\{ \beta^{1/2}, \beta^{-4}\}
\end{equation} 
gives that  $|\lambda|\gamma_j\prec 1$ for $j=2,3,4$ and then \eqref{n67} implies that $c_1(\beta,\lambda)\prec \beta^{-1} +\beta^3$. It  follows from \eqref{89} that $\varkappa_1\prec \beta^{-2} + \beta^9$. 

Consider low temperatures, $T=1/\beta\prec 1$. Then $\varkappa_1\prec \beta^9$ and so
\begin{equation}
\max\Big[ 1,\varkappa_1\kappa (1+\varkappa_1\kappa/\delta), \alpha,\varkappa_1\Big] \prec \beta^{18}\max\big[ 1,\alpha,\kappa(1+\kappa/\delta)\big].
\label{24}
\end{equation}
Also, one easily sees that $\lim_{\beta\rightarrow\infty} \|IP_\r\|$ is a fixed nonzero value, and thus 
\begin{equation}
\min\Big[ 1,a,\delta/\kappa^2, g^{3/2}\Big]\prec \min\Big[ 1,a,\delta/\kappa^2, \|I P_\r\|, g^{3/2}\Big]
\label{25}
\end{equation}
for small temperatures. Combining \eqref{24} and \eqref{25} gives the bound
\begin{equation}
T^{18} \frac{\min\big[ 1,a,\delta/\kappa^2, g^{3/2}\big]}{\max\big[ 1,\alpha,\kappa(1+\kappa/\delta)\big]}\prec \lambda_0, \mbox{\qquad for $T\prec 1$ small.}
\end{equation}
Then $|\lambda|\prec \lambda_0^{4/3}$ is satisfied provided \eqref{lambdasmall} holds. This completes the proof of Proposition \ref{prop1.2}.\qed

\subsection{Level shift operators and Fermi golden rule condition (A5)}
\label{FGRsect}

It is shown in \cite{Merkli2001} that if \eqref{n97} is satisfied,  then $L_\lambda$ has a simple eigenvalue at zero for small $\lambda\neq 0$. It then follows that $L_\lambda$ does not have any other eigenvalues, except for that single one at the origin (\cite{JP2001, BR}). This argument holds even if the spectrum of $H_\s$ is degenerate.
Denoting as in \cite{Markov1} the number of eigenvalues of $L_\lambda$ for $\lambda\neq 0$ by $m'_e$, we see that condition \eqref{n97}  implies that  $m'_0=1$ and $m'_e=0$ for all $e\neq 0$. In particular, the only invariant state for $\lambda\neq 0$ is the coupled equilibrium state.

More precisely, as shown in \cite{Merkli2001}, condition \eqref{n97} implies that the spectrum of the level shift operators $\Lambda_e$ for $e\neq 0$ lies in the open upper complex plane ${\mathbb C}_+ = \{ z\ :\ {\rm Im}z>0\}$ and that $\Lambda_0$ has a simple eigenvalue at the origin (eigenvector $\Omega_{\s,\beta}$, see \eqref{n47.1}) and all other eigenvalues of $\Lambda_0$ are in ${\mathbb C}_+$. Therefore,  $\Lambda_e$ has exactly $m'_e$ real eigenvalues, as required in condition (A5) of that paper. This characterization of the spectrum of the $\Lambda_e$ is usually called the {\em Fremi Golden Rule Condition}. 

The assumption (A5) in \cite{Markov1}  contains another part, namely that the eigenvalues of $\Lambda_e$ are simple. We have 
\begin{equation}
\Lambda_e = -P_e I P^\perp _e(L_0-e+\i 0_+)^{-1} IP_e
\label{n98}
\end{equation}
with $I$ given in \eqref{n40} and where $P_e=P_{\s,e}\otimes P_\r$, where $P_\r$ is given in \eqref{n58} and $P_{\s,e}$ is the eigenprojection of $L_\s$ associated to the eigenvalue $e$. As $I$ contains two terms, \eqref{n98} is the sum of four terms,
\begin{eqnarray}
\lefteqn{
	\Lambda_e = }\label{n99}\\
&&  -P_e \{ G\otimes \bbbone_\s\otimes \varphi_\beta(\tau_\beta g) \} (L_0-e+\i 0_+)^{-1} \{G\otimes \bbbone_\s\otimes \varphi_\beta(\tau_\beta g)\}P_e\nonumber\\
&& +P_e \{ G\otimes \bbbone_\s\otimes \varphi_\beta(\tau_\beta g) \} (L_0-e+\i 0_+)^{-1} \{\bbbone_\s\otimes {\mathcal C}G{\mathcal C}\otimes \varphi_\beta(e^{-\beta u/2}\tau_\beta g)\}P_e\nonumber\\
&&+P_e \{\bbbone_\s\otimes {\mathcal C}G{\mathcal C}\otimes \varphi_\beta(e^{-\beta u/2}\tau_\beta g)\} (L_0-e+\i 0_+)^{-1} \{ G\otimes \bbbone_\s\otimes \varphi_\beta(\tau_\beta g) \}P_e\nonumber\\
&&-P_e \{\bbbone_\s\otimes {\mathcal C}G{\mathcal C}\otimes \varphi_\beta(e^{-\beta u/2}\tau_\beta g)\} (L_0-e+\i 0_+)^{-1} \{\bbbone_\s\otimes {\mathcal C}G{\mathcal C}\otimes \varphi_\beta(e^{-\beta u/2}\tau_\beta g)\}P_e.
\nonumber
\end{eqnarray}
The partial trace over the reservoir part is calculated using the formula
\begin{eqnarray}
\lefteqn{
	P_\r \varphi_\beta(F)  (L_0-e+\i 0_+)^{-1}  \varphi_\beta(G) P_\r =\tfrac12 P_\r a_\beta(F)  (L_0-e+\i 0_+)^{-1}  a^*_\beta(G) P_\r}\nonumber\\
&& \qquad  \qquad \qquad = \tfrac12 P_\r \int_{{\mathbb R}\times S^2} \bar F(u,\Sigma) G(u,\Sigma) (L_\s-e+u+\i 0_+)^{-1}du d\Sigma,
\label{n100}
\end{eqnarray}
valid for any two functions $F,G\in L^2({\mathbb R}\times S^2)$. 
One obtains from \eqref{n99}
\begin{eqnarray}
\lefteqn{
	\Lambda_e =}\label{n101}\\
&& -\tfrac12 P_{\s,e} (G\otimes\bbbone_\s) \int_{{\mathbb R}\times S^2} \frac{\big|g(|u|,\Sigma)\big|^2}{|1-e^{-\beta u}|} (L_\s-e+u+i 0_+)^{-1}  u^2 du d\Sigma\  (G\otimes\bbbone_\s) P_{\s,e}\nonumber\\
&&+\tfrac12 P_{\s,e} (G\otimes\bbbone_\s) \int_{{\mathbb R}\times S^2} \frac{e^{-\beta u/2}\big|g(|u|,\Sigma)\big|^2}{|1-e^{-\beta u}|} (L_\s-e+u+i 0_+)^{-1}  u^2 du d\Sigma\  (\bbbone_\s\otimes{\mathcal C}G{\mathcal C}) P_{\s,e}\nonumber\\
&&+\tfrac12 P_{\s,e} (\bbbone_\s\otimes {\mathcal C}G{\mathcal C}) \int_{{\mathbb R}\times S^2} \frac{e^{-\beta u/2}\big|g(|u|,\Sigma)\big|^2}{|1-e^{-\beta u}|} (L_\s-e+u+i 0_+)^{-1}  u^2 du d\Sigma\ (G\otimes\bbbone_\s)  P_{\s,e}\nonumber\\
&&-\tfrac12 P_{\s,e} (\bbbone_\s\otimes {\mathcal C}G{\mathcal C}) \int_{{\mathbb R}\times S^2} \frac{\big|g(|u|,\Sigma)\big|^2}{|e^{\beta u}-1|} (L_\s-e+u+i 0_+)^{-1}  u^2 du d\Sigma\ (\bbbone_\s\otimes {\mathcal C}G{\mathcal C})  P_{\s,e}.\nonumber
\end{eqnarray}
The condition that $\Lambda_e$ should have simple spectrum can then be verified for concrete cases, where $g, G$ are given explicitly, see {\em e.g.} Section \ref{SBsec}. Note that if $e$ is a simple eigenvalue of $L_\s$, then $P_{\s,e}$ has rank one and so $\Lambda_e$ has automatically simple spectrum. Note also that $e=0$ is always a degenerate eigenvalue of $L_\s$.

\subsubsection{Spin-Boson model, proof of Proposition \ref{prop1.3}}
\label{SBsec}

Consider $H_\s$ and $G$ to be given as in \eqref{131.1}. The energies of $H_\s$ are $E_1=\Delta/2$, $E_2=-\Delta/2$ and so the eigenvalues of $L_\s$ are $e\in\{-\Delta, 0,\Delta\}$ with multiplicities $m_{-\Delta}=m_\Delta=1$ and $m_0=2$. The gap of $L_\s$ is thus
\begin{equation}
g=\Delta.
\label{g=d}
\end{equation}
Set $\phi_{ij}=\phi_i\otimes\phi_j\in{\mathbb C}^2\otimes{\mathbb C}^2$. We have 
\begin{equation}
P_{-\Delta} = |\phi_{21}\rangle\langle\phi_{21}|,\quad P_0 = |\phi_{11}\rangle\langle \phi_{11}| + |\phi_{22}\rangle\langle \phi_{22}|, \quad  P_{\Delta} = |\phi_{12}\rangle\langle\phi_{12}|.
\end{equation}
Using \eqref{n101} we find
\begin{eqnarray}
\Lambda_{\Delta} &=& P_{\Delta} \Big[ \ \frac{\i\pi}{2} \Delta^2 |G_{12}|^2 \coth(\beta\Delta/2) \int_{S^2} |g(\Delta,\Sigma)|^2\nonumber\\
&& \qquad +\frac{\i\pi}{2}(G_{11}-G_{22})^2 \int_{{\mathbb R}\times S^2} u^2\frac{|g(|u|,\Sigma)|^2}{|1-e^{-\beta u}|} \delta_{{\mathbb R}}(u)\, du d\Sigma\nonumber\\
&&\qquad -\tfrac12 |G_{12}|^2  \int_{{\mathbb R}\times S^2} u^2\big|g(|u|,\Sigma)\big|^2 \coth(\beta|u|/2)\, {\rm P.V.} \frac{1}{u-\Delta}\, du d\Sigma\nonumber\\
&& \qquad -\tfrac12 \int_{{\mathbb R}\times S^2} u^2\frac{\big|g(|u|,\Sigma)\big|^2}{|1-e^{-\beta u}|} (G_{11}-e^{-\beta u/2}G_{22})^2\,  {\rm P.V.}\frac{1}{u} \,dud\Sigma\ \Big]
\label{135.1}
\end{eqnarray}
where $\delta_{\mathbb R}(u)$ is the Dirac distribution acting on functions of the variable $u\in\mathbb R$ and ${\rm P.V.}$ denotes the principal value. Those distributions arise due to the identity
\begin{equation}
(u-u_0+\i 0_+)^{-1} = -\i \pi \delta_{{\mathbb R}}(u-u_0) + {\rm P.V.} \frac{1}{u-u_0},\qquad \forall u_0\in\mathbb R.
\end{equation}
The level shift operator $\Lambda_{-\Delta}$ is obtained from the expression on the right side of \eqref{135.1} by replacing $P_\Delta$ by $P_{-\Delta}$ and changing the sign of the two real terms in $[\cdots]$ (see also Theorem 1.1 of \cite{Mlso}). It is automatic that $\Lambda_0=\i \Gamma_0$ for a self-adjoint $\Gamma_0$ (again by Theorem 1.1 of \cite{Mlso}) and an explicit calculation using \eqref{n101} gives
\begin{eqnarray}
\Lambda_0 &=&\i\gamma_0
\begin{pmatrix}
e^{\beta\Delta/2} & -1\\
-1 & e^{-\beta \Delta/2}
\end{pmatrix}\\
\gamma_0 &=& \pi\Delta^2|G_{12}|^2 \frac{e^{-\beta\Delta/2}}{1-e^{-\beta \Delta}} \int_{S^2}|g(\Delta,\Sigma)|^2 d\Sigma.
\end{eqnarray}
Note that the vector $(e^{-\beta\Delta/2},1)^T$ is in the kernel of $\Lambda_0$. The nonzero eigenvalue of $\Lambda_0$ is thus its trace, $2\i |G_{12}|^2 J(\Delta)\coth(\beta\Delta/2)$. Expressed in terms of the spectral density \eqref{specdens}, we thus obtain the following eigenvalues of the level shift operators:
\begin{eqnarray}
a_0^{(0)} &=& 0\label{137}\\
a_0^{(1)} &=& 2\i |G_{12}|^2 J(\Delta)\coth(\beta\Delta/2)\\
a_{\Delta} &=& \i |G_{12}|^2 J(\Delta) \coth(\beta\Delta/2) + \i (G_{11}-G_{22})^2 \big(\frac{J(\omega)}{1-e^{-\beta\omega}}\big)\Big|_{\omega=0} \label{139}\\
&& -\tfrac12 |G_{12}|^2  \int_{{\mathbb R}\times S^2} u^2\big|g(|u|,\Sigma)\big|^2 \coth(\beta|u|/2)\, {\rm P.V.} \frac{1}{u-\Delta}\, du d\Sigma\nonumber\\
&&  -\tfrac12 \int_{{\mathbb R}\times S^2} u^2\frac{\big|g(|u|,\Sigma)\big|^2}{|1-e^{-\beta u}|} (G_{11}-e^{-\beta u/2}G_{22})^2\,  {\rm P.V.}\frac{1}{u} \,dud\Sigma\\
a_{-\Delta} &=& \overline{a_{\Delta}} 
\label{141}
\end{eqnarray}
Let us now estimate the parameters $a,\alpha,\delta,\kappa$, \eqref{FGR}-\eqref{kappa}. In the remainder of this section we write $A\prec B$ to mean that $A\le {\rm const.}B$ for a constant independent of $\lambda,a,\alpha,\delta,\kappa$ and independent of the inverse temperature $\beta$.

\medskip

We start with an estimate for $a$. The nonzero imaginary parts of the eigenvalues of $\Lambda_{\pm \Delta}$ and  $\Lambda_0$ are $|G_{12}|^2J(\Delta)\coth(\beta\Delta/2) + (G_{11}-G_{22})^2 \big( \frac{J(\omega)}{1-e^{-\beta \omega}} \big) |_{\omega=0}$ and $2|G_{12}|^2 J(\Delta)\coth(\beta\Delta/2)$, respectively. We have $J(\omega)\sim\omega^{2+2p}$  for  $\omega\sim 0$, where $p$ is the infrared parameter, see \eqref{paramvalues}. Thus $\big( \frac{J(\omega)}{1-e^{-\beta \omega}} \big) |_{\omega=0}=0$ for $p>-1/2$ and $\big( \frac{J(\omega)}{1-e^{-\beta \omega}} \big) |_{\omega=0}=c/\beta$ for $p=-1/2$. We thus get, for $G_{12}\neq 0$ and $J(\Delta)\neq 0$, 
\begin{equation}
1\prec a\prec 1\qquad \mbox{for \ $T=1/\beta \prec 1$.}
\label{142}
\end{equation}

Now we estimate $\alpha=\max_{e,s}|a_e^{(s)}|= \max\{|a_0^{(1)}|, |a_\Delta|\}$, \eqref{alpha}. From the above analysis of $a$, we know how to estimate $a_0^{(1)}$ as well as the imaginary part of $a_{\Delta}$; they are $\prec 1$.  Consider the the integral involving the principal value of $(u-\Delta)^{-1}$. The part of the integral coming from the domain $|u-\Delta|>\Delta/2$ is bounded $\prec 1$ for $1\prec \beta$. On the part where the distribution becomes singular, $|u-\Delta|\le \Delta/2$, we estimate the integral as follows. We write it as 
\begin{eqnarray}
\lefteqn{(2/\pi)\,  {\rm P.V.} \int_{|u-\Delta|\le \Delta/2} J(u)\coth(\beta u/2)\frac{1}{u-\Delta} du}\nonumber\\
&=& (2/\pi)\coth(\beta \Delta/2)\,  {\rm P.V.} \int_{|u-\Delta|\le \Delta/2} J(u)\frac{1}{u-\Delta} du\nonumber\\
&& + (2/\pi)\,   \int_{|u-\Delta|\le \Delta/2} J(u)\frac{\coth(\beta u/2) - \coth(\beta \Delta/2)}{u-\Delta} du.
\label{144}
\end{eqnarray}
In the second integral on the right side of \eqref{144} the principal value distribution has become simply the function $(u-\Delta)^{-1}$. We use the mean value theorem, 
\begin{equation}
\frac{\coth(\beta u/2) - \coth(\beta \Delta/2)}{u-\Delta} = -\frac{\beta}{2}\frac{1}{(e^{\beta\xi/2} - e^{-\beta\xi/2})^2}
\end{equation}
for some $\xi$ between $u$ and $\Delta$. Clearly, $\Delta/2\le \xi\le 3\Delta/2$ and so since $1\prec\beta$, we have $e^{\beta\Delta/2}\prec e^{\beta\xi}\prec (e^{\beta\xi/2} - e^{-\beta\xi/2})^2$. We conclude that $|\eqref{144}|\prec  \beta$.

Next we analyze the last integral in \eqref{139}. To alleviate the analysis, we shall assume that $p>-1/2$. Then the ${\rm P.V.}\frac{1}{u}$ is simply the function $u^{-1}$ and it follows readily that the integral has the upper bound $\prec 1+1/\beta\prec 1$.

We conclude that 
\begin{equation}
\alpha\prec \beta \qquad \mbox{for $T=1/\beta\prec 1$.}
\label{145}
\end{equation}

Next we estimate $\delta$, the gap between the eigenvalues of the level shift operators, \eqref{delta}. Only $\Lambda_0$ has to be taken into account, since $\Lambda_{\pm\Delta}$ are one-dimensional. We have 
\begin{equation}
1 \prec \delta=|a_0^{(1)}|\prec 1,\qquad \mbox{for \ $T=1/\beta\prec 1$.}
\label{146}
\end{equation}

Finally, we have (see \eqref{kappa})
\begin{equation}
\kappa =1,
\label{147}
\end{equation} 
since the eigenprojections of all the level shift operators are orthogonal in the case at hand. 

We are now able to simplify the condition \eqref{lambdasmall}. Using \eqref{142}, \eqref{146} and \eqref{147} we get  (recall that we do not estimate uniformly in $g=\Delta$)
\begin{equation}
1 \prec \min\big[ 1,a,\delta/\kappa^2\big] .
\label{148}
\end{equation}
Next, using \eqref{145}, \eqref{146} and \eqref{147}, we get
\begin{equation}
\max\big[ 1,\alpha,\kappa(1+\kappa/\delta)\big] \prec \beta.
\label{149}
\end{equation}
We combine \eqref{148} and \eqref{149} to obtain
\begin{equation}
\frac{1}{\beta} \prec \frac{\min\big[ 1,a,\delta/\kappa^2, g^{3/2}\big]}{\max\big[ 1,\alpha,\kappa(1+\kappa/\delta)\big]}.
\label{150}
\end{equation}
We conclude that \eqref{lambdasmall} holds provided $T\prec 1$ and 
\begin{equation}
|\lambda| \prec T^{76/3}.
\label{151}
\end{equation}
This shows the bounds of Proposition \ref{prop1.2}. It remains to calculate the  Davies generator $K$, \eqref{K}. Its expression is given in the Appendix of \cite{MAOP}. This completes the proof of Proposition \ref{prop1.2}. \qed

\medskip
{\bf Acknowledgements.\ } The author thanks two anonymous referees for their valuable input. The author was supported by a {\em Discovery Grant} from the {\em National Sciences and Engineering Research Council of Canada } (NSERC). Thanks also go to Lukas Schalleck from the journal {\em Quantum}, for thoughtfully following both this and the previous companion paper during the publication process.

\end{document}